\documentclass[preprint,10pt]{aastex} 

\pdfoutput=1

\usepackage{multirow}

\usepackage{natbib}

\usepackage[english]{babel}

\usepackage{color}

\usepackage{lscape}

\usepackage{geometry} 
\usepackage{graphicx} 

\usepackage{booktabs} 
\usepackage{array} 
\usepackage{paralist} 
\usepackage{verbatim} 
\usepackage{subfigure} 

\usepackage{fancyhdr} 
\usepackage{sectsty}
\allsectionsfont{\sffamily\mdseries\upshape} 

\title{Chemical and Physical Conditions in Molecular Cloud Core 
DC~000.4$-$19.5 (SL42) in Corona Australis\footnote{Based on 
observations collected at the European Southern Observatory, 
Chile, and the use of Herschel Science Archive data. 
\textit{Herschel} is an ESA space observatory with science instruments 
provided by European-led Principal Investigator consortia and with 
important participation from NASA.}}
\author{
E.~Hardegree-Ullman\altaffilmark{1},
J.~Harju\altaffilmark{2,3},
M.~Juvela\altaffilmark{3},
O.~Sipil$\rm{\ddot{a}}$\altaffilmark{3},
D.~C.~B.~Whittet\altaffilmark{1},
and S.~Hotzel\altaffilmark{4,5}
}

\altaffiltext{1}{New York Center for Astrobiology and Department of Physics, Applied Physics, \& Astronomy, Rensselaer Polytechnic Institute, 110 Eighth Street, Troy, NY 12180, USA; hardee@rpi.edu}
\altaffiltext{2}{Finnish Centre for Astronomy with ESO (FINCA), University of Turku, V\"ais\"al\"antie 20, FI-21500, Piikki\"o, Finland}
\altaffiltext{3}{Department of Physics, P.O. Box 64, FI-00014, University of Helsinki, Finland} 
\altaffiltext{4}{Observatory, FI-00014, University of Helsinki, Finland}
\altaffiltext{5}{GRS mbH, D-50667, Cologne, Germany}

\def\arcsec{\hbox{$^{\prime\prime}$}}
\def\arcmin{\hbox{$^{\prime}$}}
\def\deg{\hbox{$^{\circ}$}}
\def\hour{\hbox{$^{\rm h}$}}
\def\min{\hbox{$^{\rm m}$}}

\def\ceio{\hbox{$\rm{C^{18}O}$}}
\def\diaz{\hbox{$\rm{N_2H^+}$}}
\def\kms{\hbox{$\rm{km~s^{-1}}$}}
\def\n2{\hbox{$\rm{N_2}$}}
\def\h2{\hbox{$\rm{H_2}$}}
\def\av{\hbox{$A_{V}$}}
\def\hmink{\hbox{$H-K$}}
\def\jminh{\hbox{$J-H$}}
\def\um{\hbox{$\mu$m}}

\begin{document}

\renewcommand{\thefootnote}{\fnsymbol{footnote}}


\begin{abstract}

Chemical reactions in starless molecular clouds are heavily dependent on 
interactions between gas phase material and solid phase dust and ices. We have observed the 
abundance and distribution of molecular gases in the cold, starless 
core \mbox{DC~000.4$-$19.5} (SL42) in Corona Australis using data from the 
Swedish ESO Submillimeter Telescope. We present column density maps determined 
from measurements of \ceio\ ($J=2-1,1-0$) and \diaz\ ($J=1-0$) emission features. 
\textit{Herschel} data of the same region allow a direct comparison to the dust component of the 
cloud core and provide evidence for gas phase depletion of CO at the highest extinctions. 
The dust color temperature in the core calculated from \textit{Herschel} maps ranges from 
roughly 10.7 to 14.0~K. 
This range agrees with the previous determinations from \textit{Infrared Space Observatory} and 
\textit{Planck} observations. The column density profile of the core can be fitted with a 
Plummer-like density distribution approaching $n(r) \sim r^{-2}$ at large distances. The core 
structure deviates clearly from a critical Bonnor-Ebert sphere. Instead, the core appears to be 
gravitationally bound and to lack thermal and 
turbulent support against the pressure of the surrounding low-density material: 
it may therefore be in the process of slow contraction. We 
test two chemical models and find that a steady-state depletion model 
agrees with the observed \ceio\ column density 
profile and the observed $N(\ceio)$ versus \av\ relationship.

{\it Key words:} dust, extinction --- infrared: ISM --- ISM: clouds --- ISM: individual objects 
(\mbox{DC~000.4$-$19.5/SL42}) --- ISM: molecules --- radio lines: ISM

\end{abstract}

\maketitle

\thispagestyle{myheadings}
\markboth{}{\bf To appear in The Astrophysical Journal, 763:45, 2013 Jan 20}


\renewcommand{\thefootnote}{\arabic{footnote}}\setcounter{footnote}{6}


\section{Introduction}

Nascent solar systems 
inherit their chemical inventories from precursor molecular cloud cores. 
Within these cores, molecules containing the abundant CHON group of
elements are of greatest importance to astrochemistry, particularly in
the context of the possible emergence of habitable planets and life. CO
is a dominant species within this group, because of both its abundance
and the vital role it plays in chemical pathways leading to complex
organic molecules (\citealt{1982A&A...114..245T}; see \citealt{2011ApJ...742...28W} and
references therein for additional discussion).

Gravitationally bound starless cores, referred to as ``prestellar" cores 
\citep{2007prpl.conf...17D,2007prpl.conf...33W}, represent sites of imminent future star 
and solar-system formation. In this paper, we investigate the interplay of molecular gas and dust in the 
starless core \mbox{DC~000.4$-$19.5} associated with the Corona Australis (CrA) 
molecular cloud complex. 
\mbox{DC~000.4$-$19.5} has been identified in many surveys of the region and possesses numerous 
alternative designations: Cloud 42/S42/SLDN42 \citep{1976A&A....53..179S}, Cloud B 
\citep{1978AJ.....83..234R}, Condensation C \citep{1996A&AS..116...21A}, 
Core 5 \citep{1999PASJ...51..911Y}, and Condensation CoA7 \citep{2000ApJ...532.1038V}, 
for example. Henceforth, the core will be referred to as SL42. 
SL42 is an isolated molecular cloud located at \mbox{R.A.\ 19\hour10\min16\fs3},
\mbox{Decl.\ $-$37\deg08\arcmin37\arcsec}, J2000.0, at a distance of 130 pc 
\citep{2008hsf2.book..735N}. No embedded young stellar objects (YSOs) 
have been detected within 5\arcmin\ of SL42, 
corresponding to the approximate radius of its core; H$\alpha$~16, the nearest YSO 
\citep{1981AJ.....86...62M}, is a weak-line T Tauri star \citep{1998A&AS..128..561B, 
2002MNRAS.336..197G} 7\farcm4 from the center of SL42. It is possible that the 
molecular cloud core which gave rise to H$\alpha$~16 was originally part 
of SL42 as evidenced by the present-day extension of the SL42 envelope toward 
H$\alpha$~16. The quiescent nature of SL42 restricts chemical reactions within the 
cloud to low temperature gas phase molecular 
interactions and surface reactions on dust grains.


Section \ref{sec:data} presents the data collection and reduction techniques for both radio 
and infrared observations of SL42. Section \ref{sec:analysis} describes the production of \ceio, 
\diaz, and \h2\ maps, the determination of visual extinction within SL42, and the estimates of 
the cloud's mass and dynamical state. Section \ref{sec:modeling} addresses the attempts to 
match chemical models to the observed properties of SL42, and Section \ref{sec:discussion} 
discusses the implications of our analysis. Finally, Section \ref{sec:summary} summarizes our findings. 


\section{Data}
\label{sec:data}

\subsection{SEST Observations}

\begin{figure}[ht]
\centering
\includegraphics[angle=0,scale=.75]{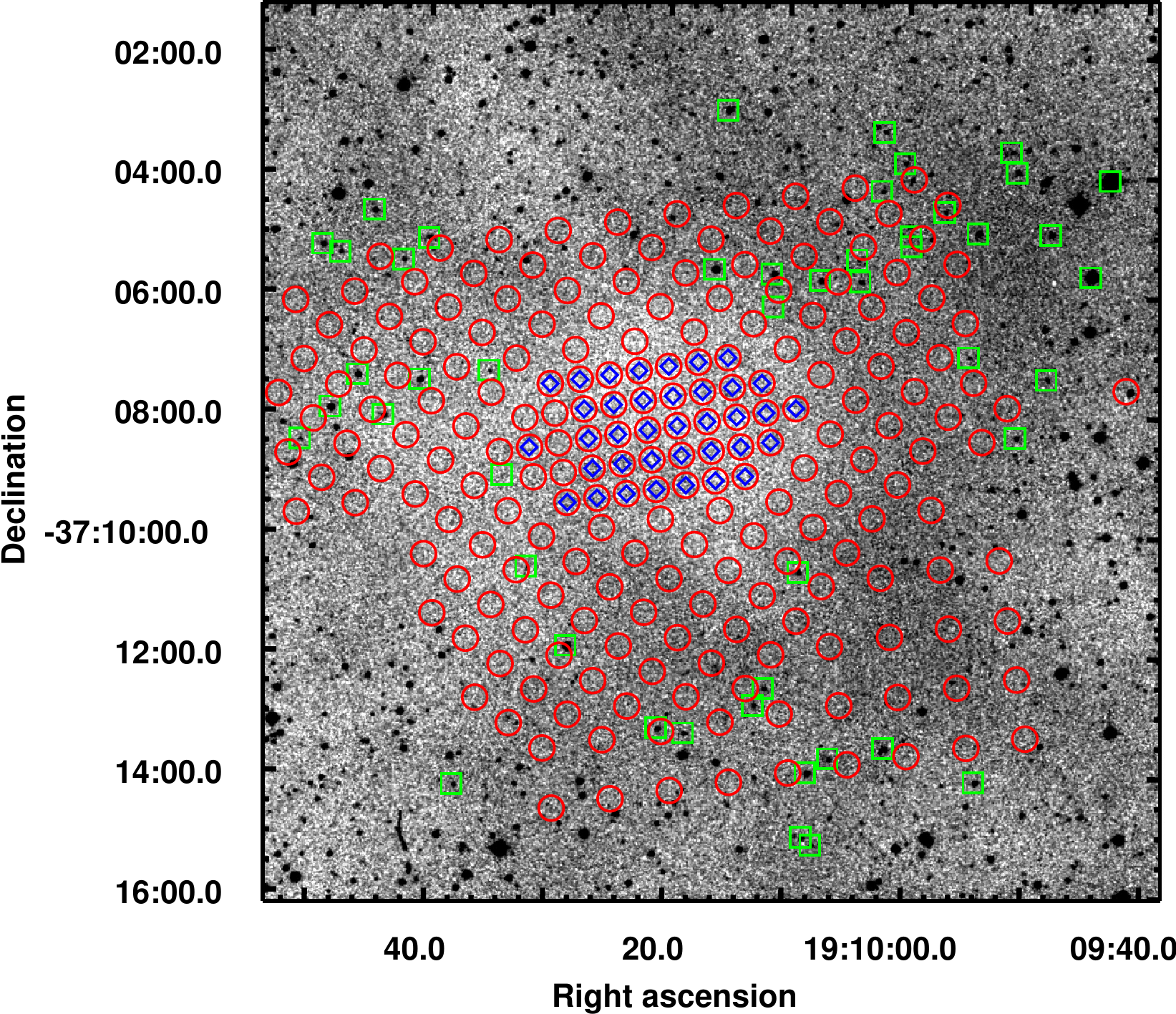}
\caption{\small Data coverage of the sample over a negative infrared DSS 
image of SL42. The red field is the distribution of \ceio\ measurements, and the size of the circles corresponds to the HPBW of the antenna at 220 GHz (25\arcsec). The blue diamonds show the
distribution of \diaz\ measurements. Locations where the observational noise exceeded the emission 
signals (i.e., S/N$<$1 or S$<$0) are not shown. The green squares indicate the locations of stars with good quality 2MASS $JHK$ photometry which met the condition $\hmink>0.4$. The image center is located at \mbox{R.A.\ 19\hour10\min16\fs3}, \mbox{Decl.\ $-$37\deg08\arcmin37\arcsec},
J2000.0. The image is $\rm{15\arcmin \times 15\arcmin}$.}
\label{fig:datacover}
\end{figure}

Radio data were collected using the 15 m Swedish ESO Submillimeter
Telescope (SEST) in 2003 March.
The molecular lines observed were \ceio\ $J=1-0$ (109.8~GHz),
\ceio\ $J=2-1$ (219.6~GHz), and \diaz\ $J=1-0$ (93.2~GHz). A region of
about $12\arcmin \times 8\arcmin$, centered on the ISOSS 
source 19102$-$3708 (\mbox{R.A.\ 19\hour10\min16\fs3}, 
\mbox{Decl.\ $-$37\deg08\arcmin37\arcsec}, J2000.0) 
was mapped in the two \ceio\ transitions simultaneously with the 3 and 
1.3~mm superconductor-insulator-superconductor receivers, using a grid 
spacing of 40\arcsec. A region of $\sim$~3\arcmin~$\times$~3\arcmin\ around the
\ceio\ emission peak was thereafter mapped in \diaz\ $J=1-0$ and
\ceio\ $J=2-1$ using a 20\arcsec\ spacing. Figure \ref{fig:datacover} shows 
the coverage of the data field over a negative infrared Digitized Sky 
Survey (DSS)\footnote{{\tt http://archive.stsci.edu/dss/acknowledging.html}, 
\copyright Anglo-Australian Observatory (AAO)/Royal Observatory, 
Edinburgh (ROE)} image of SL42.

All observations were performed in the frequency switching mode using
an integration time of one minute per position. The two mixers used at
the same time were connected to a 2000 channel acousto-optical
spectrometer (AOS) which was split in two bands of 43 MHz each. The AOS
channel width corresponds to 0.14$\:\kms$ at 93~GHz, 0.12$\:\kms$ at 
110~GHz, and 0.06$\:\kms$ at 220~GHz. The system temperatures were around
240, 260, and 450~K at 93, 110, and 220~GHz, respectively. The rms
noise level attained was typically $\sim$~0.1~K at 3~mm and $\sim$~0.2~K at 1.3~mm.

The half-power beam width (HPBW) of the antenna was 47\arcsec\ at 110~GHz 
and 25\arcsec\ at 220~GHz. The pointing and focus were checked at
3$-$4 hr intervals toward circumstellar SiO maser line sources, and
we estimate that the pointing accuracy was typically about 3\arcsec.

Calibration was done by the chopper wheel method. To convert the
observed antenna temperatures ($T_{ A}^{*}$) to the radiation
temperatures ($T_{R}$) the former were divided by the assumed
source-beam coupling efficiencies ($\eta_{C}$) for which we adopted
the main beam efficiencies ($\eta_{MB}$) of the telescope
interpolated to the frequencies used, i.e., 0.73, 0.71, and 0.61 at
93, 110, and 220~GHz, respectively.

\subsection{SEST Data Reduction}

\begin{figure}
\centering
\vskip -1in
\includegraphics[angle=0, width=\textwidth]{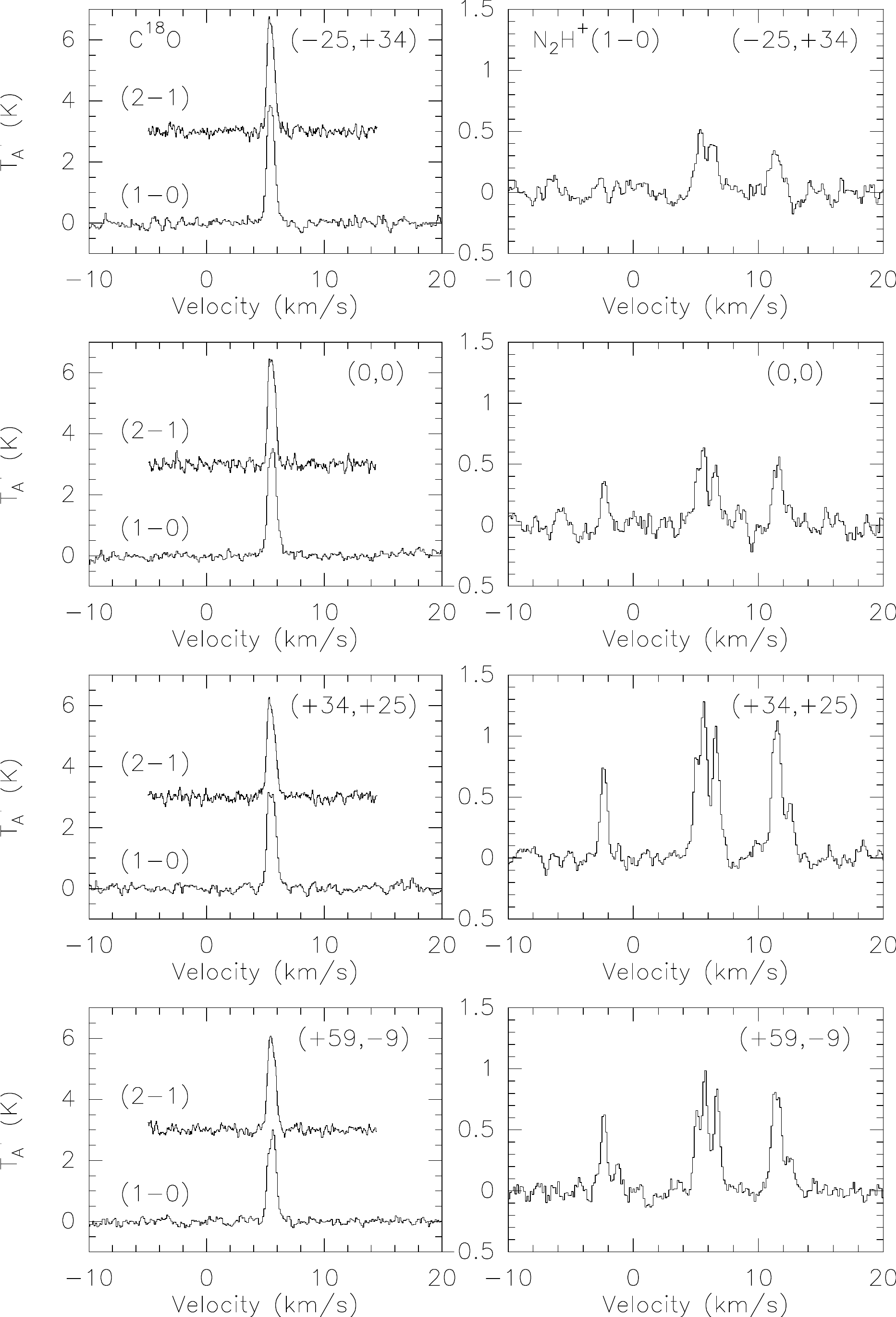}
\caption{\small Example spectra from the sample. The left column shows \ceio\ $J=1-0$ and
$J=2-1$ emission lines. The right column shows \diaz\ $J=1-0$ lines. The four
positions represent the brightest spots in \ceio\ (top two rows) and \diaz\ (bottom two rows). 
The multiple peaks in \diaz\ spectra are due to hyperfine splitting. The origin position for the SEST 
observations, $(0\arcsec, 0\arcsec)$, is located at \mbox{R.A.\ 19\hour10\min16\fs3}, 
\mbox{Decl.\ $-$37\deg08\arcmin37\arcsec}, J2000.0.}
\label{fig:spec}
\end{figure}

The spectra were reduced using the Continuum and Line Analysis 
Single-dish Software (CLASS) program of the Grenoble Image and Line Data Analysis 
Software (GILDAS) package for the Institut de Radioastronomie Millim$\acute{e}$trique 
(IRAM)\footnote{\tt
 http://www.iram.fr/IRAMFR/GILDAS}. (See Figure \ref{fig:spec} 
for example spectra.) The reduction involved the
subtraction of the baseline using a polynomial fit and fitting
Gaussian profiles to the detected lines to calculate the peak antenna
temperatures ($T_{A}^*$), local standard of rest (LSR) velocities ($v_{\rm{LSR}}$), and the
full-width half-max (FWHM) line widths ($\Delta v$). The intensities were converted to the
main-beam brightness temperature scale by $T_{MB} = T_{A}^{*}/\eta_{MB}$, 
where $\eta_{MB}$ is the main-beam
efficiency. Assuming uniform beam filling, $T_{MB}$ represents the
brightness temperature $T_{B}$.
 
The \diaz\ $J=1-0$ lines contain seven hyperfine (hf) components.
To fit the hf structure we used the frequencies and the relative line
strengths given in \citet{1995ApJ...455L..77C}. Besides $v_{\rm{LSR}}$ and
$\Delta v$, the hf fit also gives the total optical depth
($\tau_{\rm{tot}}$) of the line which is the sum of peak optical
thicknesses of the seven components. The fit can be used to calculate the
optical depth profile ($\tau_{v}$) of the line (as a function of
radial velocity $v$). The \ceio\ lines do not have hf structure,
and the lines could be well fitted with a single Gaussian. The lines
are likely to be optically thin in SL42, a relatively isolated, small cloud.

The \ceio\ and \diaz\ column densities were determined from the
reduced spectra by assuming that the rotational levels are populated
according to the local thermodynamic equilibrium (LTE). For
\ceio\ it was furthermore assumed that the lines are optically
thin.

When the excitation temperature ($T_{ex}$) and the
integrated optical depth ($\int\tau_{v}dv$) of the observed 
transition $u \rightarrow l$ are known, the column density of the 
molecules in the upper level ($N_{u}$) can be calculated from the formula
\begin{equation}
N_{u}=\frac{8\pi}{\lambda^3}\frac{1}{A_{ul}} F(T_{ul},T_{ex})\int\tau_{v}dv,
\end{equation}
where $A_{ul}$ is the Einstein coefficient for spontaneous
emission, $\lambda= c/\nu_{ul}$ is the wavelength of the
transition, $T_{ul}\equiv h\nu_{ul}/k_{B}$, and the
function $F(T_{ul},T)$ is defined by
\begin{equation}
F(T_{ul},T) \equiv \frac{1}{e^{T_{ul}/T}-1}.
\end{equation}

For an optically thin line, as in the case of \ceio, the integral 
$\int\tau_{v}dv$ can be estimated from the integrated intensity using the
antenna equation:
\begin{equation}
\int T_{B}(v)dv\approx T_{ul}[F(T_{ul},T_{ex})-F(T_{ul},T_{bg})]\int\tau_{v}dv.
\end{equation}

For the derivation of the \ceio\ excitation temperature, the
\ceio\ $J=2-1$ spectral map was convolved to the same angular
resolution as the $J=1-0$ map, and the $T_{ex}$ was calculated
from integrated intensity ratio of the two lines using the antenna
equation.

The integrated $\tau$ of the \diaz\ line was obtained from hf fit 
results assuming a Gaussian profile: 
\begin{equation}
 \int\tau_{v}dv=\frac{\sqrt{\pi}}{2\sqrt{\ln{2}}}\Delta v\tau_{\rm{tot}}.
\end{equation}
The $T_{ex}$ of \diaz\ was calculated by substituting 
the $T_{B}$ and $\tau$ values at the line peak into the antenna
equation:
\begin{equation}
T_{B}(v)=T_{ul}[F(T_{ul},T_{ex})-F(T_{ul},T_{bg})](1 - e^{-\tau_{v}}).
\end{equation}

The total column densities ($N_{\rm{tot}}$) including all rotational levels
were estimated according to the LTE assumption: 
\begin{equation}
N_{\rm{tot}}=\frac{N_u}{g_u}e^{E_u/kT_{ex}}Z(T_{ex}), 
\end{equation} 
where $Z(T_{ex})$ is the rotational partition function at the
temperature $T_{ex}$, and $E_u$ is the energy of the level $u$
with respect to the rotational ground level.


\subsection{\textit{Herschel} Data}
\label{sec:herschel}

The $15\arcmin\times15\arcmin$ \textit{Herschel} maps of SL42 used in this study 
were extracted from the extensive
mapping of the CrA region collected by the Spectral 
and Photometric Imaging Receiver (SPIRE) and Photodetector 
Array Camera and Spectrometer (PACS) instruments as part of the Herschel Gould Belt Survey
(\citealt{2010A&A...518L.102A}; see also {\tt\small{http://www.herschel.fr/cea/gouldbelt/en/}}). 
An overview of the \textit{Herschel Space Observatory} is given in \citet{2010A&A...518L...1P}. 
The SPIRE and PACS instruments are described in \citet{2010A&A...518L...3G} and
 \citet{2010A&A...518L...2P}, respectively.

The pipeline-reduced data from the SPIRE+PACS parallel mode survey of CrA are publicly available at the Herschel Science Archive\footnote{\tt http://herschel.esac.esa.int/Science\_Archive.shtml}. The original CrA maps cover a region of about $5\degr\times2\degr$ at the wavelengths 500, 350, and 250 \um\ (SPIRE), and 160 and 70 \um\ (PACS). The FWHM values of the Gaussians fitted to the beam profiles are, in order of decreasing wavelength, 37$\arcsec$, 25$\arcsec$, 18$\arcsec$, 12$\arcsec\times$16$\arcsec$, and 6$\arcsec\times$12$\arcsec$, the PACS beams being clearly elongated in observations carried out in the parallel mode.

In order to calculate the distributions of the dust temperature ($T_{d}$) and 
the optical depth ($\tau_{\lambda}$) of the dust
emission, the \textit{Herschel} maps at 500, 350, 250, and 160 \um\ were first convolved
with a Gaussian beam to a resolution of 40\arcsec\ (FWHM), and the intensity
distributions were fitted pixel by pixel with a modified blackbody
function, $I_\nu\approx B_\nu(T_{d})\tau_\nu\propto B_\nu(T_{d})\nu^{\beta}$, 
which characterizes optically thin thermal dust emission at far-IR and 
submillimeter wavelengths. The observed values of $\beta$, the spectral index, are typically close to 2.0 (e.g.,
\citealt{1996A&A...312..256B,2010ApJ...708..127S,2011A&A...527A.111J}), but
higher values have been reported for cold regions (e.g., \citealt{2003A&A...404L..11D,
2008A&A...481..411D,2011A&A...536A..23P}). The recent study of the Taurus-Auriga molecular
cloud found an average value of $\sim$~1.8 but again significant
anticorrelation between dust temperature and the spectral index. In our analysis 
we fixed the dust opacity spectral index to a constant value of $\beta=2.0$, which is 
used in several \textit{Herschel} studies so the $T_{d}$ and $N(\h2)$ maps presented in this work are 
directly comparable with previous results. This is consistent with the Planck Early Cold Core (ECC)
estimate for SL42, $\beta=2.4\pm0.6$ (see Section~\ref{sec:other}), especially considering that
the Planck value is the estimate for the cold dust component rather
than for the total dust emission from this source. The 
pipeline-calibrated in-beam flux densities were color corrected to
monochromatic flux densities at the standard wavelengths in the course
of $T_{d}$ fitting using the adopted shape of the source spectrum
(a modified blackbody spectrum with $\beta=2.0$) and the spectral
response functions of SPIRE and PACS photometers for an extended
source\footnote{See the SPIRE and PACS Observer's Manuals at {\tt http://herschel.esac.esa.int/Documentation.shtml}}. In
addition, the SPIRE pipeline in-beam flux densities were converted from point
source calibration to extended source calibration before the
temperature fitting and color correction.

\begin{figure} 
\centering
	\begin{subfigure}
		\centering
		\includegraphics[width=0.45\textwidth]{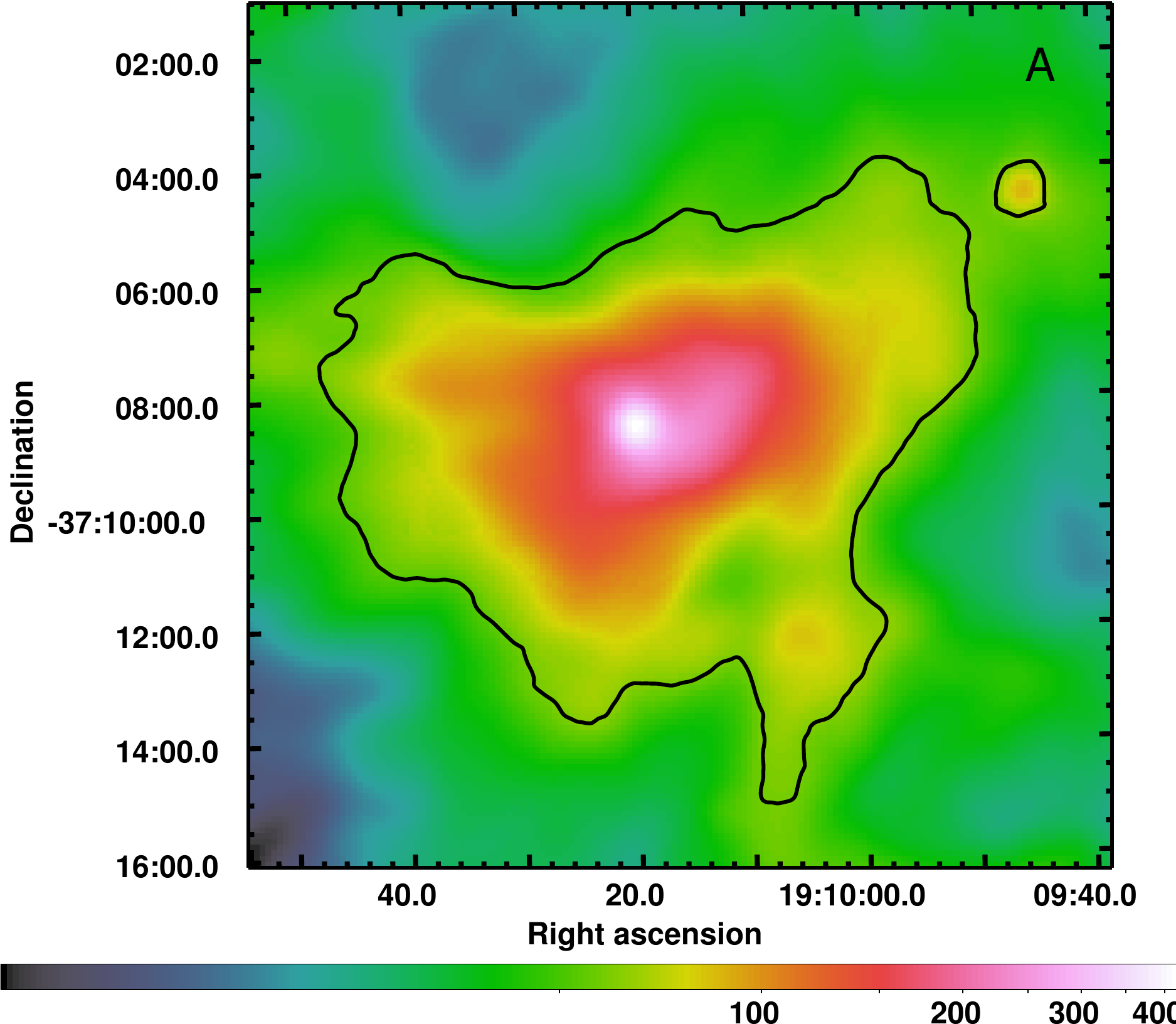}
	\end{subfigure}
	\begin{subfigure}
		\centering
		\includegraphics[width=0.45\textwidth]{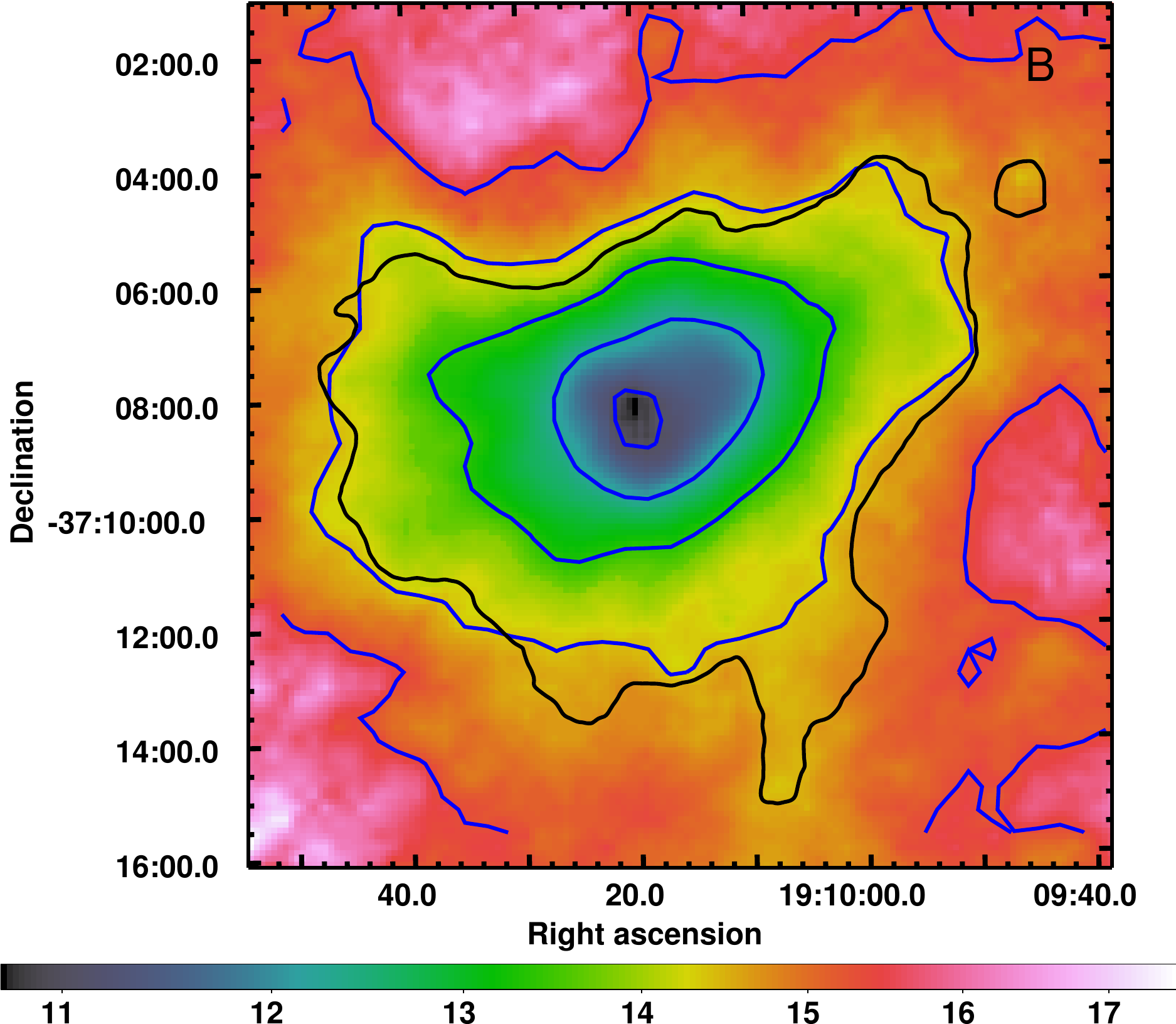}
	\end{subfigure}
	\begin{subfigure}
		\centering
		\includegraphics[width=0.45\textwidth]{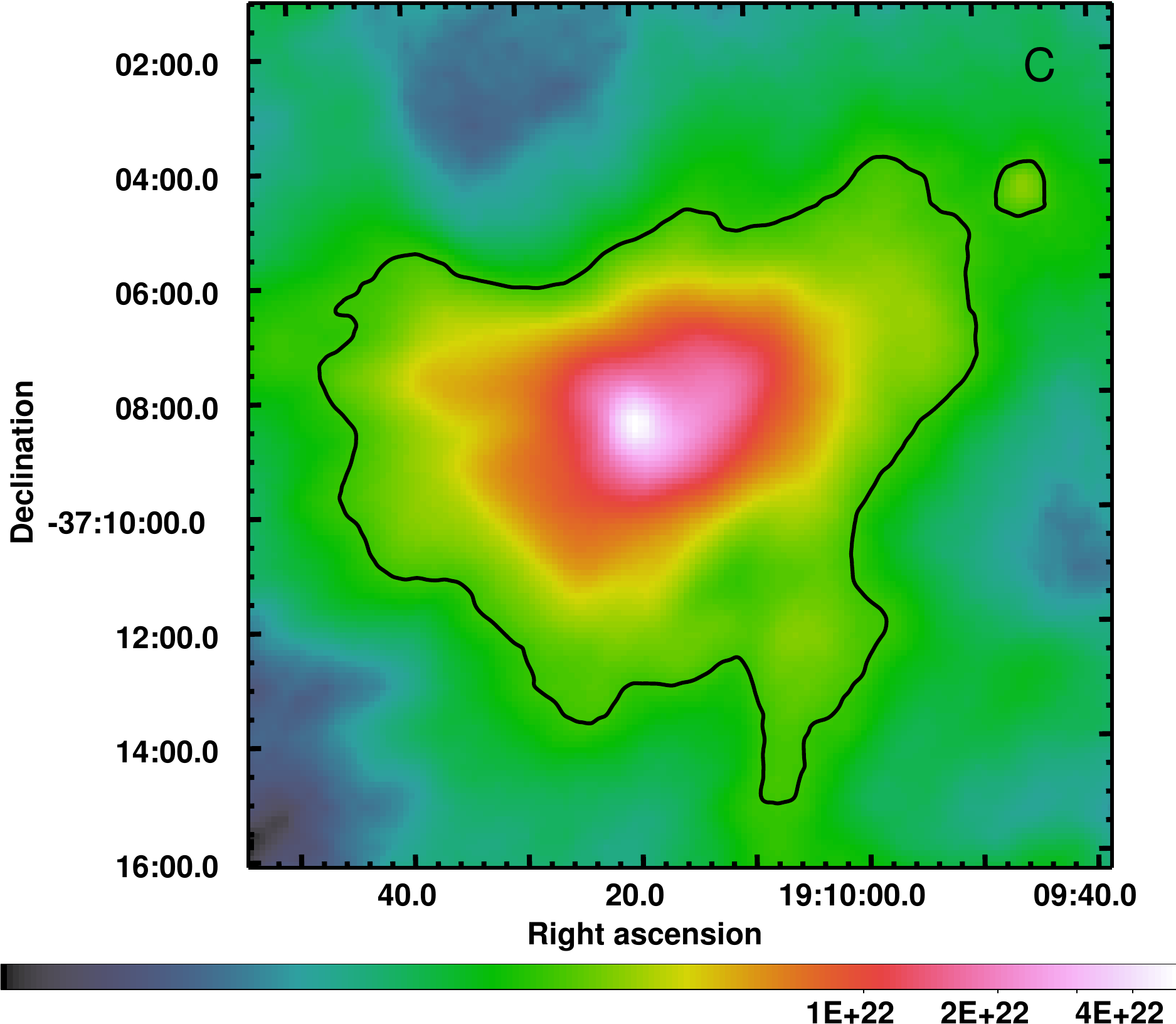}
	\end{subfigure}
\caption{Colorbar units are MJy~sr$^{-1}$, K, and cm$^{-2}$, respectively. (A) \textit{Herschel} intensity map of thermal dust emission at
the wavelength $\lambda=250~\um$ in a $15\arcmin\times15\arcmin$ 
region around SL42. (B) Dust
temperature map of the same region, created by fitting a modified blackbody function to the \textit{Herschel}
intensity maps at 500, 350, 250, and 160 \um\ (see Section~\ref{sec:herschel}). Blue contours range 
from 11~K at the center to 15~K in increments of 1~K. (C) \h2\ column density map determined from the
far-infrared optical depth (see Equation~(\ref{eq:nh2})). The black contour shows where 
$N(\h2)=3.3\times10^{21}~\rm{cm}^{-2}$.}
\label{figure:Herschel250}
\end{figure}

The resulting intensity map ($I_{250\mu\rm{m}}$) at $\lambda$=250 \um\ 
and the $T_{d}$ map are shown in Figure \ref{figure:Herschel250}.
The optical depth map ($\tau_{250\mu\rm{m}}$), calculated from
$\tau_{250\mu\rm{m}} = I_{250\mu\rm{m}}/B_{250\mu\rm{m}}(T_{d})$, is proportional to the molecular hydrogen column density ($N(\h2)$). The column density can be obtained from 
\begin{equation}
\label{eq:nh2}
N(\h2) = \frac{\tau_{250\mu\rm{m}}}{\kappa_{250\mu\rm{m}} {\bar m}_{\rm{H_2}}},
\end{equation} 
where $\kappa$ is the dust ``opacity", or absorption cross-section
per unit mass of {\sl gas}, and ${\bar m}_{\rm{H_2}}$ ($=2.8$ amu $=4.65\times10^{-24}$ g) is
the average mass of gas {\sl per $H_2$ molecule} (assuming
10\% He). The \h2\ column density errors were estimated by a Monte Carlo method using the 
$1\sigma$ error maps provided for the three SPIRE bands and a 7\% uncertainty of the 
absolute calibration for all four bands according to the information given in SPIRE and 
PACS manuals. Different realizations of the $N(\h2)$ map were calculated by 
combining the four intensity maps with the corresponding error maps, assuming 
that the error in each pixel is normally distributed. The $N(\h2)$ error in each 
pixel was obtained from the standard deviation of one thousand realizations. 
For the dust opacity at 250 \um, we used the value 
$\kappa_{250\mu\rm{m}}^{\rm g} = 0.085~{\rm cm^{2}~g^{-1}}$ which is a factor 
of 1.17 lower than the value $\kappa_{250\mu\rm{m}}^{\rm g} = 0.1~{\rm cm^{2}~g^{-1}}$ 
consistent with the parameter $C_{250}~(=1/\kappa_{250\mu\rm{m}})$ given in Table 1 of
\citet{1983QJRAS..24..267H}. This choice was motivated by the comparison between \h2\ column 
densities from \textit{Herschel} and visual extinctions from 2MASS as described in Section \ref{sec:av}. 




\subsection{ISOPHOT and \textit{Planck} Observations of Infrared Continuum Emission}
\label{sec:other}

The Infrared Space Observatory photo-polarimeter (ISOPHOT) Serendipity 
Survey (ISOSS) recorded the 170 \um\ sky
brightness when the satellite was slewing between two pointed
observations. SL42 is one of the ``cold cores" \citep{2000A&A...364..769T} 
detected in this survey. The dust temperature calculated
by correlating the ISOPHOT 170 \um\ and 
\textit{Infrared Astronomical Satellite} (IRAS)/IRAS Sky Survey Atlas (ISSA) 100 \um\ 
intensities is $T_{d} = 11.1^{+1.8}_{-0.8}$ K. A detailed description of
the method can be found in \citet{2001A&A...372..302H}. The ISOPHOT detector
had four pixels with slightly different tracks. The width of the slew was
about $3^\prime$, and the effective angular resolution in the scan
direction was $2\farcm2$. The surface brightness distribution
recorded by the two pixels which passed near the cloud center shows a
Gaussian bump. It has an FWHM of $5\farcm7$ and a maximum intensity of 52~MJy~$\rm{sr}^{-1}$. 
Assuming spherical symmetry, the total 170~\um\ flux density of the core is 162~Jy. 
It sits on top of a 15~MJy~$\rm{sr}^{-1}$ pedestal, which is likely to correspond to the 
low-density envelope around the core. Its position was used as the origin position for 
the SEST observations. 

SL42 can be found in the Planck Early Release Compact Source Catalogue 
(ERCSC)\footnote{See {\tt http://irsa.ipac.caltech.edu/applications/planck/}} as the object 
PLCKECC G000.37$-$19.51. The core is detected at the Planck frequencies from 100 to 857 GHz 
($\lambda=3.0-0.35$ mm). 
The coordinates of the 857 GHz maximum (PLCKERC857 G000.36$-$19.49) 
are at R.A.\ and Decl.\ offsets $(51\arcsec, -9\arcsec)$ from the SEST origin. The dust 
temperature and the emissivity index in the
cold core are $T_{d} = 10.2\pm0.6$ K and $\beta=2.4\pm 0.6$,
respectively. These are calculated from the 857, 545, and 353 GHz fluxes,
after the subtraction of the warm component (see the Explanatory Supplement\footnote{{\tt irsa.ipac.caltech.edu/data/Planck/release/ercsc\_v1.3/explanatory\_supplement\_v1.3.pdf}}). 
The ERCSC gives furthermore a total flux density $S_{857 \rm{ GHz}}=271\pm9$ Jy and an 
angular size of $6\farcm9\times6\farcm1$ (FWHM) for the core. The \textit{Planck} beam (FWHM) is 
$5\farcm1\times3\farcm9$ at 857 GHz. The beam size is similar at other submillimeter 
wavelengths.


\section{Analysis}
\label{sec:analysis}

\subsection{\ceio\ and \diaz\ Maps, Line Characteristics, and Column Densities}

\begin{figure} 
\centering
	\begin{subfigure}
		\centering
		\includegraphics[width=0.7\textwidth]{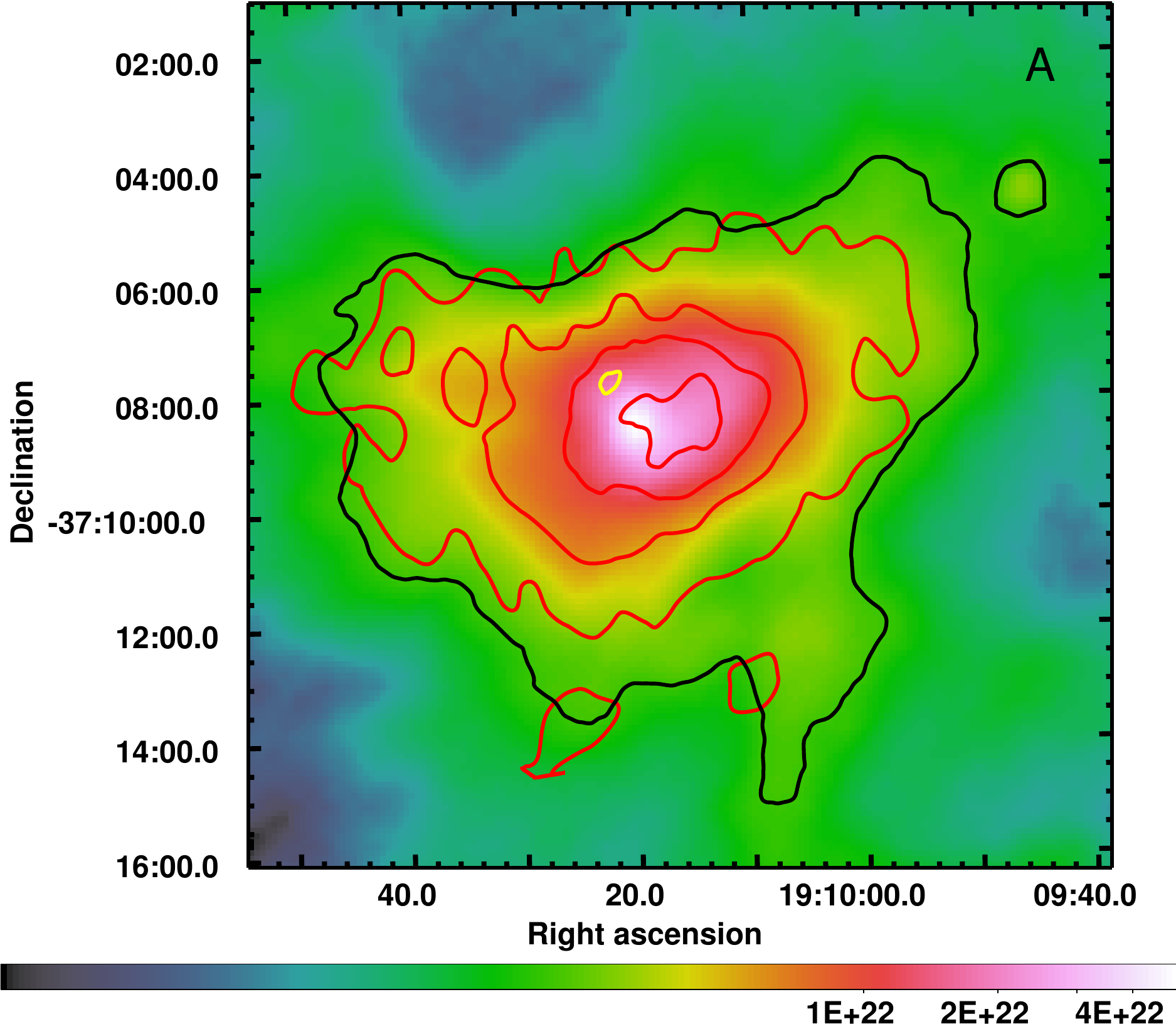}
	\end{subfigure}
	\begin{subfigure}
		\centering
		\includegraphics[width=0.7\textwidth]{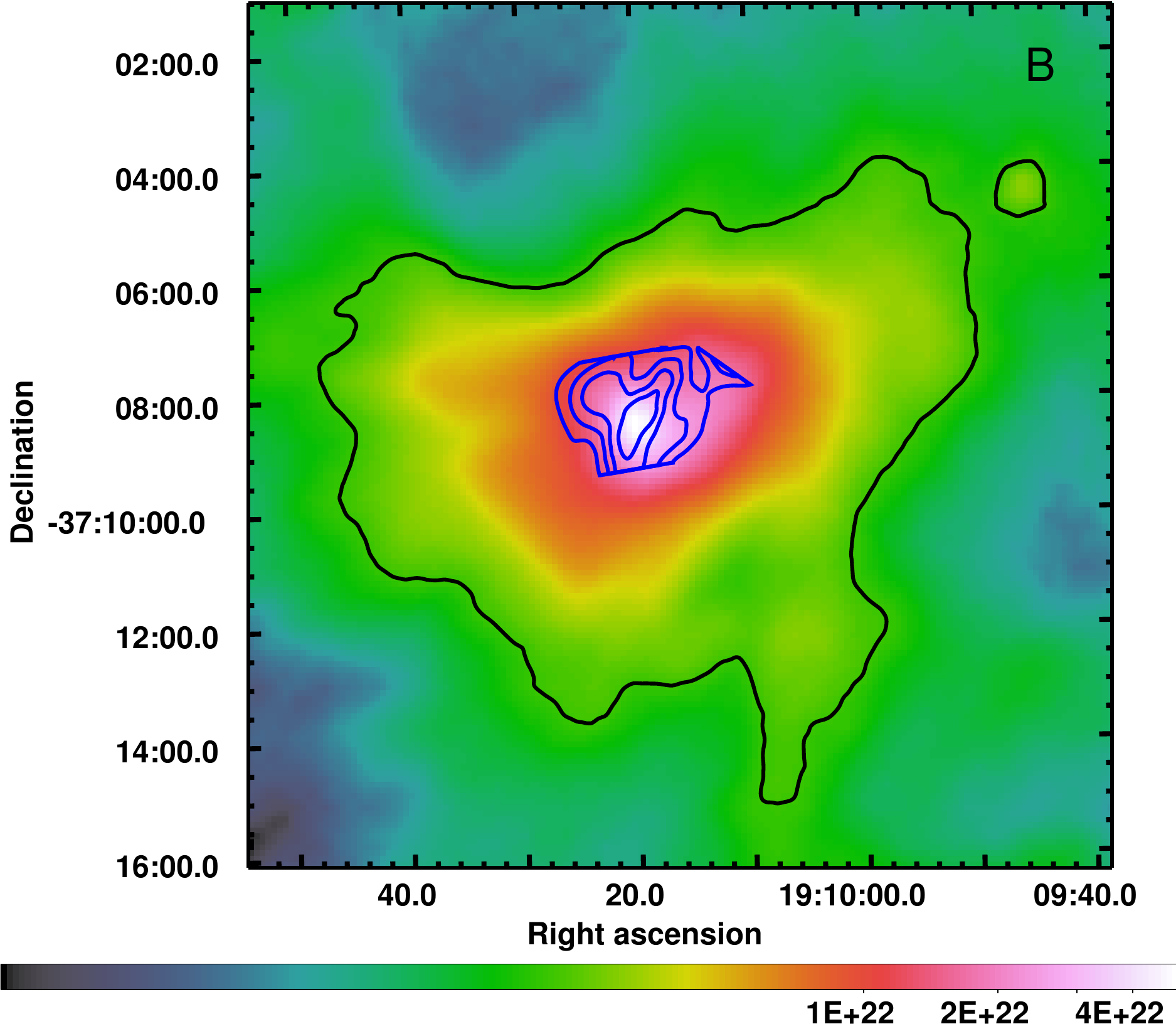}
	\end{subfigure}
\caption{Contour maps of (A) \ceio\ and (B) \diaz\ superimposed on the \h2 column 
density map of SL42. The contours for \ceio\ (red) range from 1 to 
4 $\rm{(10^{15}~cm^{-2})}$ in increments of 1; yellow indicates a depression 
contour. The contours for \diaz\ (blue) range from 2 to 11 $\rm{(10^{12}~cm^{-2})}$ in 
increments of 3. The maximum \ceio\ detection is located 
at (0\arcsec, 0\arcsec), and the maximum \diaz\ detection is located at (34\arcsec, 25\arcsec) 
giving a spatial offset of $\sim$~42\arcsec~($\sim$~5500 AU at a distance of 130 pc). 
The black contour shows where $N\rm{(\h2)=3.3\times10^{21}~cm^{-2}}$. The colorbars are in units 
of cm$^{-2}$.}
\label{fig:contour}
\end{figure}

Figure \ref{fig:contour} shows the column density maps of the
\ceio\ $J=2-1$ and \diaz\ $J=1-0$ lines. 
A maximum \ceio\ detection of 4.53 $\pm$ 0.09 ($\rm{10^{15}\ cm^{-2}}$) occurs at 
positional offsets of $(0\arcsec, 0\arcsec)$, and a maximum \diaz\ detection of 13.40 
$\pm$ 1.60 ($\rm{10^{12}\ cm^{-2}}$) occurs at positional offsets of $(34\arcsec, 25\arcsec)$. 
The \ceio\ $J=2-1$ and \diaz\ $J=1-0$ line parameters and total
$N(\ceio)$ and $N(\diaz)$ toward both the mentioned maxima and the
second brightest positions in each molecule are listed in
Table~\ref{table:lineparameters}. The spectra from these four
positions are shown in Figure \ref{fig:spec}.


\subsection{Visual Extinction (\av)}
\label{sec:av}

Dust extincts light from background sources via absorption and scattering to make them 
appear redder and dimmer. We used two methods in order to estimate the amount of visual 
extinction, and thereby dust, in SL42. First, visual extinctions (\av) were determined using 
Two Micron All Sky Survey (2MASS) $JHK$ photometry as described 
in \citet{2008ApJS..176..457S} for field stars toward SL42 (marked in green in Figure 
\ref{fig:datacover}). 

\begin{figure}[ht]
	\centering
	\includegraphics[angle=90,scale=.55]{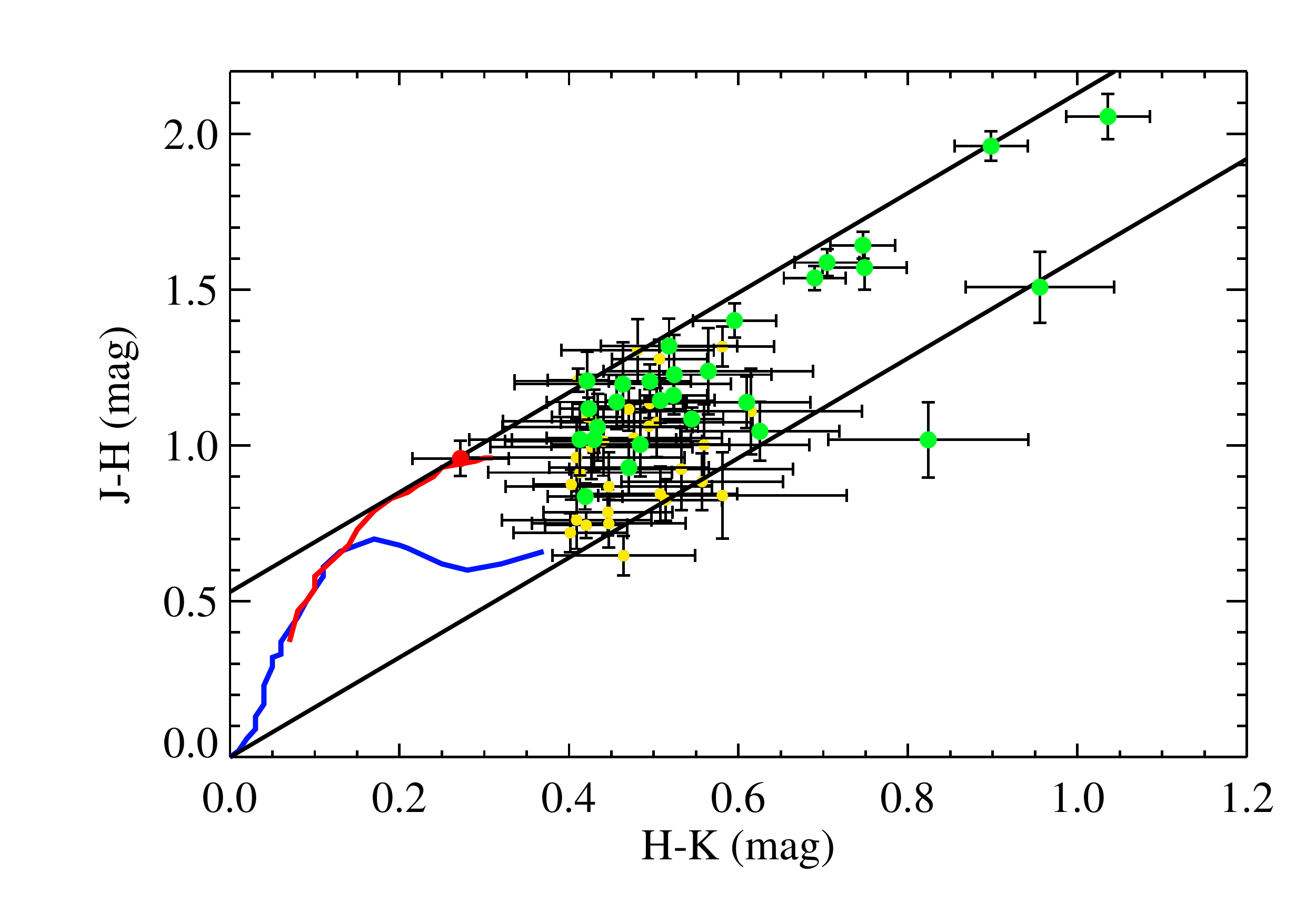}
\caption{\small Color-color diagram of all reddened ($\hmink>0.4$) field stars 
behind SL42 (equivalent to the green squares in Figure~\ref{fig:datacover}.) 
Green circles denote stars within the SEST data field, while yellow circles denote 
stars beyond the SEST field. H$\alpha$~16 is shown as a red point below the $H-K$ 
cutoff. The reddening slope for SL42 is 1.6. Red and blue mark the intrinsic color lines for 
giants and dwarfs, respectively.}
	\label{fig:jhk}
	\end{figure}

Figure \ref{fig:jhk} shows 
the color-color diagram for all of the field stars employed in this \av\ survey. We 
included only those stars from the catalog without flags indicating contamination or poor 
photometric quality of the observations. To remove unreddened or anomalous stars from 
the sample, the following color constraints were applied:
\begin{equation}
\mathrm{1.6(\hmink)+0.0<(\jminh)<1.6(\hmink)+0.53}
\end{equation}
\begin{equation}
\mathrm{0.4<(\hmink)}.
\end{equation}
The value of the reddening vector for the entire CrA cloud is 1.6, but the slope 
in the $J-H$ versus $H-K$ diagram does not change when only the SL42 region is 
considered. Stars outside of the parallel lines in Figure \ref{fig:jhk} were excluded 
since they could not be dereddened to the intrinsic color lines. 

It should be noted that the visual extinctions calculated from the 2MASS color-color diagram are 
conservative estimates. They were calculated using the minimum possible color excesses 
($E(J-K)$) necessary to place the reddened field stars on the intrinsic color lines. Most of the 
reddening vectors of the field stars in our sample cross the intrinsic color lines in two places, once 
in a region of late spectral type and once in a region of early spectral type. It is likely that 
most of our field stars are of late spectral type, so the minimum extinction values we have assumed 
are correct. Even if this is not the case for all of the field stars, the maximum extinction values 
are higher at most by a factor of two, and accounting for a few early type stars should not significantly 
change the trend seen in Figure~\ref{fig:avco}, detailed in Section \ref{sec:discussion}. 

Figure \ref{fig:jhk} presents no clear evidence for previously unknown YSOs in SL42. This is consistent 
with the \textit{Herschel} data, which confirm that SL42 is a starless core because the dust temperature decreases 
toward the center of the cloud. H$\alpha$~16 appears in Figure \ref{fig:jhk} as the red point below our 
$H-K$ color cutoff. \citet{1992ApJ...397..520W} classify H$\alpha$~16 as a class II YSO by its spectral energy distribution (SED).

The 2MASS method for estimating extinction is considered reliable because it 
measures its effects directly without making assumptions about the emissive 
properties of the intervening dust. However, this method is only 
useful in the outer regions of SL42 because the opacity in the core is so high that 
field stars become undetectable. In order to sample the cloud's extinction more 
completely, we used the canonical relationship 
$N(\h2)/\av = 9.4\times10^{20}$ cm$^{-2}$ mag$^{-1}$ 
\citep{1978ApJ...224..132B, 2012ApJ...751...28M} as a second method to obtain dust 
extinction values using \textit{Herschel} data. The calculation of $N(\h2)$ from the \textit{Herschel} 
intensity maps of SL42 depends inversely on the assumed dust opacity (see Equation
(\ref{eq:nh2})); therefore, it is important that the correct value of kappa is chosen. The 
2MASS photometry provides an independent measurement of extinction at the edge of 
the cloud, so we adjusted $\kappa_{250\mu\rm{m}}^{\rm g}$ such that $N(\h2)$ estimates 
from \textit{Herschel} 
converted to \av\ overlap the 2MASS data at low extinctions. This scaling gives a value 
of $\kappa_{250\mu\rm{m}}^{\rm g} = 0.085~{\rm cm^{2}~g^{-1}}$. 


\subsection{Mass and Dynamical State}
\label{section:MassDynamics}

The \h2\ column density map created from \textit{Herschel} data as described
in Section~\ref{sec:herschel} can be used to estimate the cloud mass. By choosing the
region above $N({\h2}) = 3.3\times10^{21}$ cm$^{-2}$ (corresponding to
$\av=3.5$ when the relationship $N(\h2)/\av =9.4\times10^{20}$ cm$^{-2}$ mag$^{-1}$ 
of \citet{1978ApJ...224..132B} is adopted) we obtain a mass of $21~M_\odot$ at the 
assumed distance of 130 pc.

The derivation of the gravitational potential energy requires an
assumption about the density distribution in the cloud. The \h2\ 
column density map has a sharp maximum at the offsets (46\arcsec, 11\arcsec), 
\mbox{R.A.\ 19\hour10\min20\fs2}, \mbox{Decl.\ $-$37\deg08\arcmin26\arcsec}, J2000.0, 
from the origin of our molecular
line maps. We calculated the azimuthally averaged \h2\ column density
profile using this position as the cloud center. The result of
the calculation is shown in Figure~\ref{fig:avprofile}. 

We also made an attempt to fit the azimuthally averaged column density
distribution with a ``Plummer-type'' \citep{2001ApJ...547..317W, 1911MNRAS..71..460P} 
outwardly decreasing number density profile using the following functional form:
\begin{equation}
n(r) = \frac{n_0}{\left(1+\frac{r^2}{r_0^2}\right)^{\eta/2}}~\mbox{when}~r\leq R_{\rm{out}},
\end{equation}
where $n(r)$ is the number density of $\h2$ as a function of radius $r$, 
$n_0$ is the gas number in the cloud center, $r_0$ is the
``flat radius" representing the radial distance where the density
increase toward the center levels off, and $R_{\rm{out}}$ is the radius
inside which this density profile is supposed to be valid. We assumed
that the cloud envelope outside this radius has a constant $\h2$
column density that forms a pedestal on top of which we see
the column density enhancement owing to the cloud core. 

The best fit to the data is obtained with the following parameter
values: $\eta=2.0$, $n_0=1.2\times10^6$ cm$^{-3}$, $r_0=15\farcs0$, and
$R_{\rm{out}} = 300\arcsec$. The exponent value $\eta=2$ means that the
density distribution approaches $n(r) \propto r^{-2}$ for $r \gg r_0$.
The average column density far from the cloud center,
$N(\h2)=3.3\times10^{21}$ cm$^{-2}$ (corresponding to $\av = 3.5$)
is used as the ``pedestal'' added to $N(\h2)$ obtained from the
density model. The predicted column density profile (convolved with a 
Gaussian to 40\arcsec\ resolution to allow comparison with the observed one) is
indicated as a dashed line in Figure~\ref{fig:avprofile}. The core
mass calculated from the adopted $\h2$ density profile up to $r=R_{\rm{out}}$ 
is $M_{\rm{core}} = 16~M_\odot$. The $5~M_\odot$ difference
between the model mass and the cloud mass from the $N(\h2)$ map is
accounted for by the ``pedestal'' with $N(\h2)=3.3\times10^{21}$
cm$^{-2}$.

The agreement between the observed (averaged) and predicted $N(\h2)$
distributions is reasonably good, and we used the model density
distribution to calculate the gravitational potential energy of the
core ($E_{\rm{grav}}$) from the integral
\begin{equation} 
E_{\rm{grav}}=-16\pi^2G{{\bar m}_{\rm{H_2}}}^2\int_0^{R_{\rm{out}}}rn(r)dr\int_0^{r}y^{2}n(y)dy=-1.0\times10^{37}~\rm{J}, 
\end{equation}
where $G$ is the gravitational constant and ${\bar m}_{\rm{H_2}}$ ($=2.8$ amu $=4.65\times10^{-24}$ g) 
is the average mass of gas {\sl per $H_2$ molecule} (assuming 10\%
He, $n(r) = n_{\rm{H_2}}(r)$). In the kinetic energy estimate we
assumed the cloud gas is isothermal with $T_{\rm{kin}}=10$ K and has a
constant non-thermal velocity dispersion, $\sigma_{\rm{N.T.}} =260$
m~s$^{-1}$. The latter value is calculated from the \ceio\ spectra in the
core region. Using these values we obtain
\begin{equation}
E_{\rm{kin}}=\frac{3}{2}M_{\rm{core}}[c_{s}^2+\sigma_{\rm{N.T.}}^2]=5.0\times10^{36}~\rm{J},
\end{equation}
where 
\begin{equation}
c_{s}\equiv\sqrt{k_{B}T_{\rm{kin}}/{\bar m}}
\end{equation} 
is the isothermal sound speed, ${\bar m}=2.33$ amu is the average particle
mass of the gas. We see that according to the estimates above, the
core satisfies the equation, $E_{\rm{grav}} + 2E_{\rm{kin}} = 0$, which in the 
absence of external pressure would indicate virial equilibrium. However, our
cloud model also includes a low density envelope which should be
exerting pressure on the core, so it seems the cloud is lacking sufficient 
thermal and/or turbulent support against collapse.\footnote{In the presence of 
an external pressure, the condition of
virial equilibrium applying to a system within a surface $S$ can be
written as $E_{\rm{grav}} + 2E_{\rm{kin}} - 3P_{\rm{ext}}V= 0$, where $P_{\rm{ext}}$ is the pressure on
the surface $S$, and $V$ is the volume of the system. A Bonnor-Ebert
sphere always satisfies this condition.}

The estimation of the kinetic and potential energies involves some 
uncertainties. The gas kinetic temperature is not directly measured, 
but we use the average dust temperature. Also, the \h2\ column densities from 
thermal dust emission are calculated using color 
temperatures and a constant dust opacity. The systematic errors 
related to these approximations are difficult to estimate. The color 
temperature is larger than the true temperature in the core center 
\citep{2012A&A...547A..11N, 2012A&A...539A..71J, 2011A&A...530A.101M}, 
and using the former temperatures we probably 
underestimate the \h2\ column densities and the mass. On the other 
hand, varying the temperature by 1~K and the velocity 
dispersion by one spectrometer channel changes the estimate of kinetic 
energy by 30 percent. Nevertheless, it is remarkable that our best 
effort estimates suggest a balance $2E_{\rm{kin}}/|E_{\rm{grav}}| \sim 1$, showing that the 
core is gravitationally bound \citep{2007ARA&A..45..565M}.


\section{Modeling}
\label{sec:modeling}

We have attempted to reproduce the observed \ceio\ and \diaz\ line profiles
(Figure~\ref{fig:spec}) and the associated column densities 
(Table~\ref{table:lineparameters}) by calculating simulated
radial abundance profiles for \ceio\ and \diaz\ using a chemical model and by using
the abundance profiles as input for a radiative transfer program. We now discuss the modeling
process and present the results.

\subsection{Chemical Modeling}
\label{sec:chem_model}

To simulate chemical evolution in the core, we divided the Plummer sphere
discussed in Section \ref{section:MassDynamics} into 40 concentric shells. 
Chemical evolution was
then calculated separately in each shell, yielding radial abundance profiles
as a function of time --- the chemical model (and the general modeling procedure)
used here is the same as the one discussed in detail in \citet{2012A&A...543A..38S}. 
The gas phase chemical reactions are adopted from the OSU reaction
file osu\_03\_2008\footnote{See http://www.physics.ohio-state.edu/$\sim$eric/}.
The initial gas phase abundances and the surface
reaction set (i.e., activation energies for select reactions) are
adopted from \citet{2010A&A...522A..42S}. 
For the chemical calculations, the visual extinction was calculated from the density
profile as a function of core radius assuming $\av=3$ at the
edge of the model core. In the calculations, it is assumed that all grains are 0.1 \um\ 
in radius and that the cosmic ionization rate is $\zeta = 1.3\times10^{-17}~\rm{s}^{-1}$.

The upper left panel in Figure~\ref{figure:chem} shows the radial abundance profiles of \ceio\ and \diaz\ (with respect to total hydrogen density $n_{\rm{H}}$ at three different time steps). Evidently the two species evolve very differently. \ceio\ is strongly depleted at the core center at later times but
retains a fairly constant abundance at the core edge as the core grows older. \diaz\ on the other hand
increases greatly in abundance throughout the core as time progresses. The upper right panel of 
Figure~\ref{figure:chem} shows the ratio of the column densities of \diaz\ and \ceio\ through the center of the model core as a function of time; the growth of the \diaz\ abundance with respect to the \ceio\ abundance as the core grows older is especially prominent here.

The \ceio\ column density map of the spherically symmetric model cloud
has a ring-like maximum with a radius of $\sim$~40\arcsec\ (5000
AU). In the convolved (40\arcsec\ resolution) image, the slight
depression at the core center has about 10\% lower column density than
the annular maximum. The corresponding \diaz\ map shows a sharp peak toward
the core center. The widths (FWHM) of the \ceio\ and \diaz\ distributions
are 200\arcsec\ and 60\arcsec, respectively. The model can thus
qualitatively explain the different extents of the \ceio\ and \diaz\ 
emissions and the fact that these molecules can peak at different
locations.

The observed \ceio\ maximum does not lie, however, symmetrically around
the $N(\h2)$ maximum but on its northwestern side. Possibly this can be
understood in terms of the asymmetric \h2\ column density
distribution which has a shoulder on the side where \ceio\ peaks, and
decreases more rapidly in other directions (see Figure \ref{fig:contour}(A)). 

\subsection{Spectral Line Modeling}
\label{sec:spec_line_model}

We attempted to reproduce the observed spectra toward the \diaz\ peak by calculating
simulated \diaz\ $J=1-0$ and \ceio\ $J=2-1$ line emission profiles with a Monte Carlo
radiative transfer program \citep{1997A&A...322..943J}, using the
chemical modeling results as input. The simulated spectra are highly
dependent on the radial abundances and hence on the column densities of
\diaz\ and \ceio, which are both variable functions of time
(Figure~\ref{figure:chem}).

The column density ratio of \ceio\ and \diaz\ toward the (34\arcsec, 25\arcsec) position 
(cf. Table~\ref{table:lineparameters}) corresponds roughly to $t\sim 4-5\times10^4$~years
in the model. The lower left and lower right panels of Figure~\ref{figure:chem} show the
simulated \ceio\ $J=2-1$ and \diaz\ $J=1-0$ line profiles, respectively, at
$t\sim 4.5\times10^4$~years. The observed peak intensities (cf.
Table~\ref{table:lineparameters}) are reproduced rather well by the model at this time step. However,
the optical depths of the model lines are much smaller than the observed
thicknesses. This is tied to the column densities predicted by the model; while the
\diaz\ and \ceio\ column density ratio in the model corresponds to the observed ratio at
$t \sim 4.5 \times 10^4$~years, the actual column densities for both species are at this time
clearly smaller than observed. This suggests that the observed column densities correspond to 
a later stage in the model.

To compare the observed and modeled column densities across the cloud we have plotted these in 
Figure~\ref{figure:model} as functions of the
visual extinction \av, calculated from \textit{Herschel} (using 
$N(\h2)/\av = 9.4\times10^{20}~\rm{cm}^{-2}~\rm{mag}^{-1}$). The model distribution corresponds 
to $t \sim 4.5 \times 10^4$~years. It is evident that the observed \ceio\ column density rises 
much more steeply with increasing \av\ than predicted by the model. The model either undervalues the 
CO production through gas-phase reactions and desorption or uses 
too high estimates for the CO destruction rates (through photodissociation and accretion) in
low-density gas. Alternatively, the discrepancy is caused by the fact
that the static model adopted here does not take account of the
increase of the density in case the core is contracting. We return to
this issue in Section~\ref{sec:discuss_modeling} 


\section{Discussion}
\label{sec:discussion}

\subsection{CO Depletion}

CO is expected to become depleted from the gas phase by adsorption onto
grains on timescales less than the expected lifetimes of dense cores
(e.g., \citealt{2010ApJ...720..259W} and references therein). In a detailed study of 
the Taurus dark-cloud complex, \citet{2010ApJ...720..259W} compare gas phase and 
solid phase CO in lines of sight toward background field stars with known 
extinction values from 2MASS data. They find that (1)~the total CO (gas + 
solid) column density agrees well with the general relation $N = 2.0
\times 10^{14}(\av - 2.0)~{\rm cm}^{-2}$ proposed by \citet{2006A&A...447..597K};
 and (2)~depletion begins to become significant at extinctions
$\av \ga 5-10$ in the Taurus cloud. 

\begin{figure}[ht]
	\centering
	\includegraphics[angle=0,scale=.7]{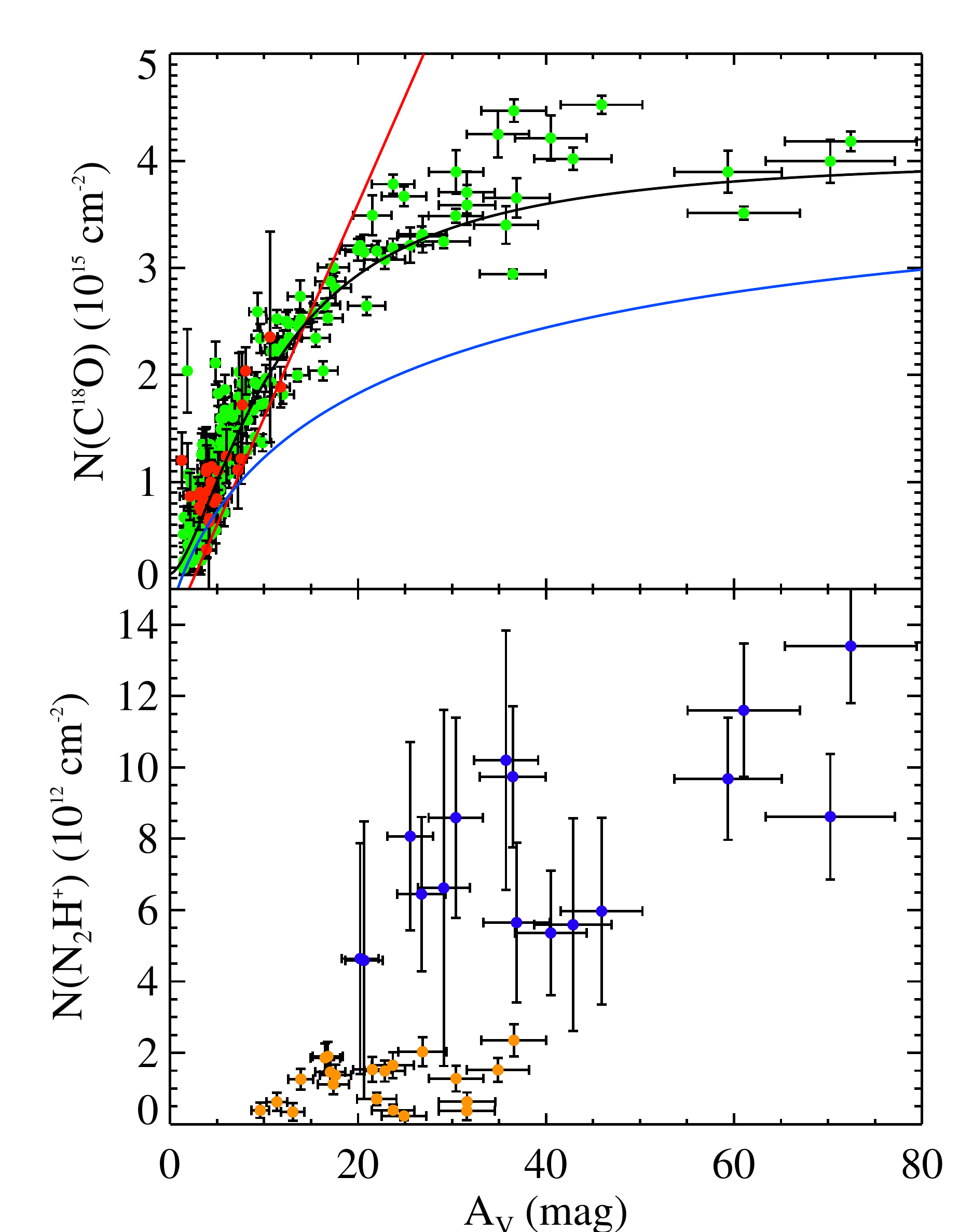}
	\caption{Top: \small \ceio\ column density as a function of extinction. 
	Red points: Interpolated $N(\ceio)$ and field star extinctions from 2MASS $JHK$ photometry.
	Green points: SEST $N(\ceio)$ and extinctions from \textit{Herschel} maps using 
	$N(\h2)/\av = 9.4\times10^{20} \rm{cm^{-2}~mag^{-1}}$ from \protect\citet{1978ApJ...224..132B}.
	Blue line: sigmoidal fit to data from the Taurus cloud (see \protect\citealt{2010ApJ...720..259W} for details). 
	Red line: average fit based on observations of molecular clouds with low depletion 
	($N=2.0 \times 10^{14} \rm{(\av\ - 2.0)~cm^{-2}}$) \protect\citep{2006A&A...447..597K}. 
	Black line: sigmoidal fit to SL42 data ($N=4.07953+(0.14037-4.07953)/(1+(\av/11.27966)^{1.55722})~\rm{cm^{-2}}$). Bottom: 
	SEST $N(\diaz)$ and extinctions from \textit{Herschel} maps. Blue points: $N(\diaz)$ found by fitting the hyperfine structure of 
	the spectral feature. Orange points: $N(\diaz)$ found assuming optically thin conditions.}
	\label{fig:avco}
	\end{figure}

\begin{figure}[ht]
	\centering
	\includegraphics[angle=0,scale=.8]{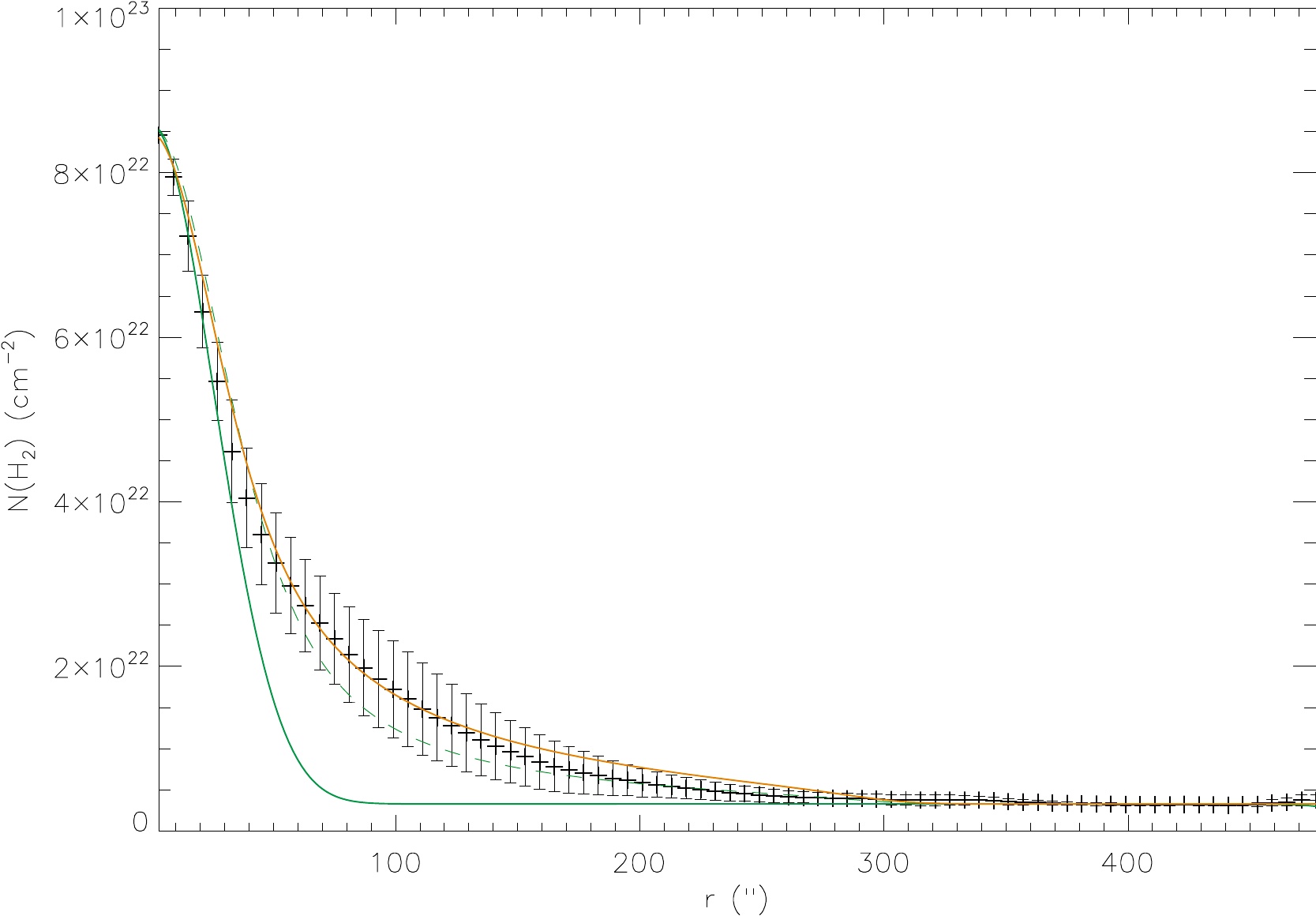}
	\caption{Azimuthally averaged \h2\ column density distribution of 
SL42 as a function of radial distance in arcseconds from the maximum 
(crosses), which occurs at \mbox{R.A.\ 19\hour10\min20\fs2}, 
\mbox{Decl.\ $-$37\deg08\arcmin26\arcsec}, J2000.0, or offsets from the SEST 
origin of (46\arcsec, 11\arcsec). The solid orange curve shows 
the \h2\ column density calculated 
from a spherically symmetric density profile following the function 
$n(r) = n_0/(1+(r/r_0)^2)$, where $n_0=1.2\times10^6$~cm$^{-3}$, 
and $r_0=15\arcsec$. The solid green curve represents a critically stable 
Bonnor-Ebert sphere ($\xi_{\rm out} = 6.45$) with the maximum 
column density agreeing with the observed one. The observed 
distribution can be fitted fairly well also with a highly super-critical 
($\xi_{\rm out}\sim 25$) Bonnor-Ebert sphere (dashed green curve).}
	\label{fig:avprofile}
	\end{figure}

\begin{figure}
\centering
\includegraphics[angle=0, scale=.6]{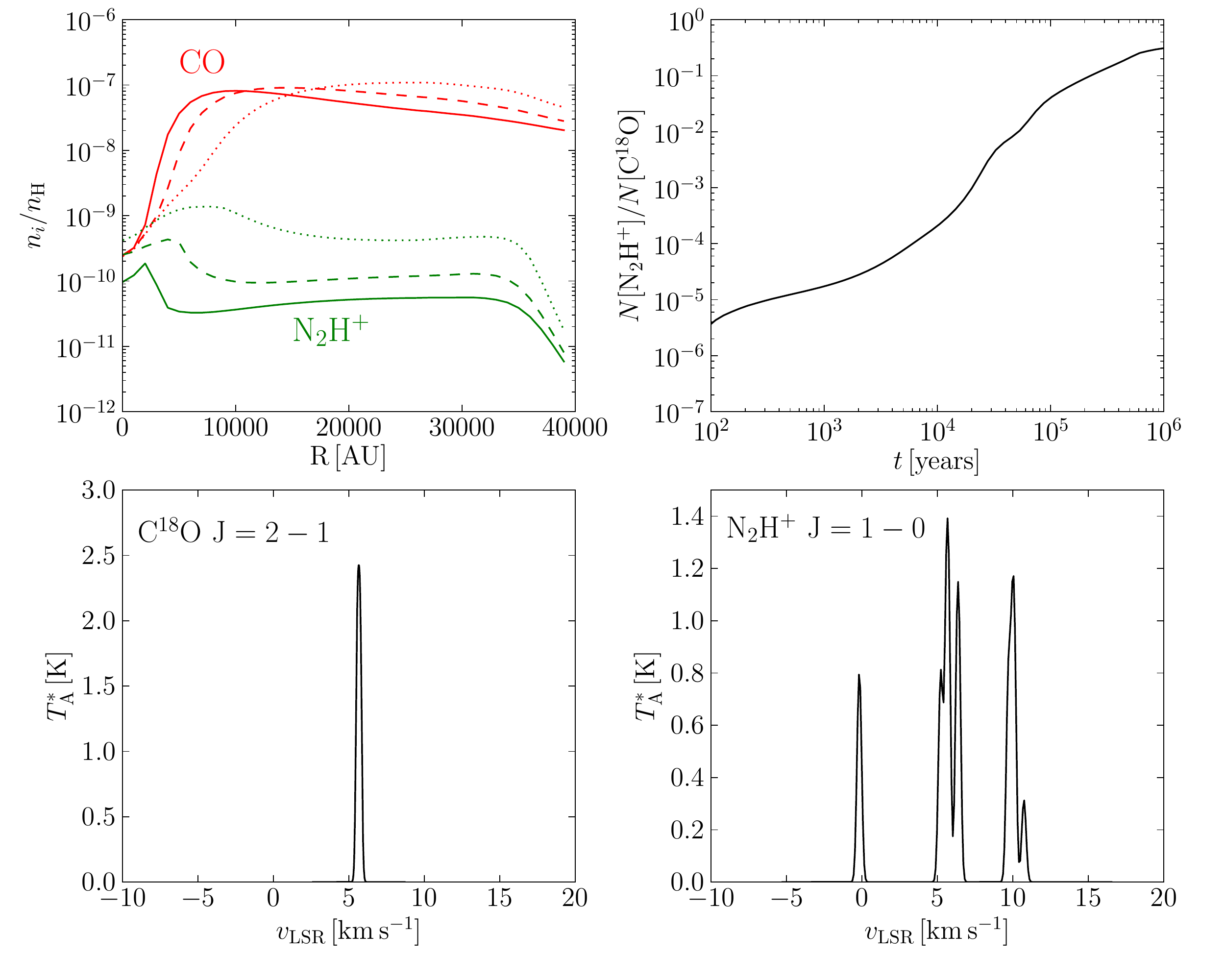}

\caption{Upper left: \ceio\ (red) and \diaz\ (green)
 radial abundances with respect to total hydrogen density $n_{\rm{H}}$ in the chemical model at
 $t = 4\times10^4$ years (solid lines), $t = 8\times10^4$ years (dashed lines) and
 $t = 2\times10^5$ years (dotted lines).
 Upper right: the ratio of \diaz\ and \ceio\ column densities through the
 center of the core model as a function of time.
 Lower left:
 the simulated \ceio\ $J=2-1$ line emission profile at $t=4.5\times10^4$~years in the antenna temperature scale.
 Lower right:
 the simulated \diaz\ $J=1-0$ line emission profile at $t=4.5\times10^4$~years in the antenna temperature scale. }
\label{figure:chem}
\end{figure}

\begin{figure}
\centering
\includegraphics[angle=0, scale=.90]{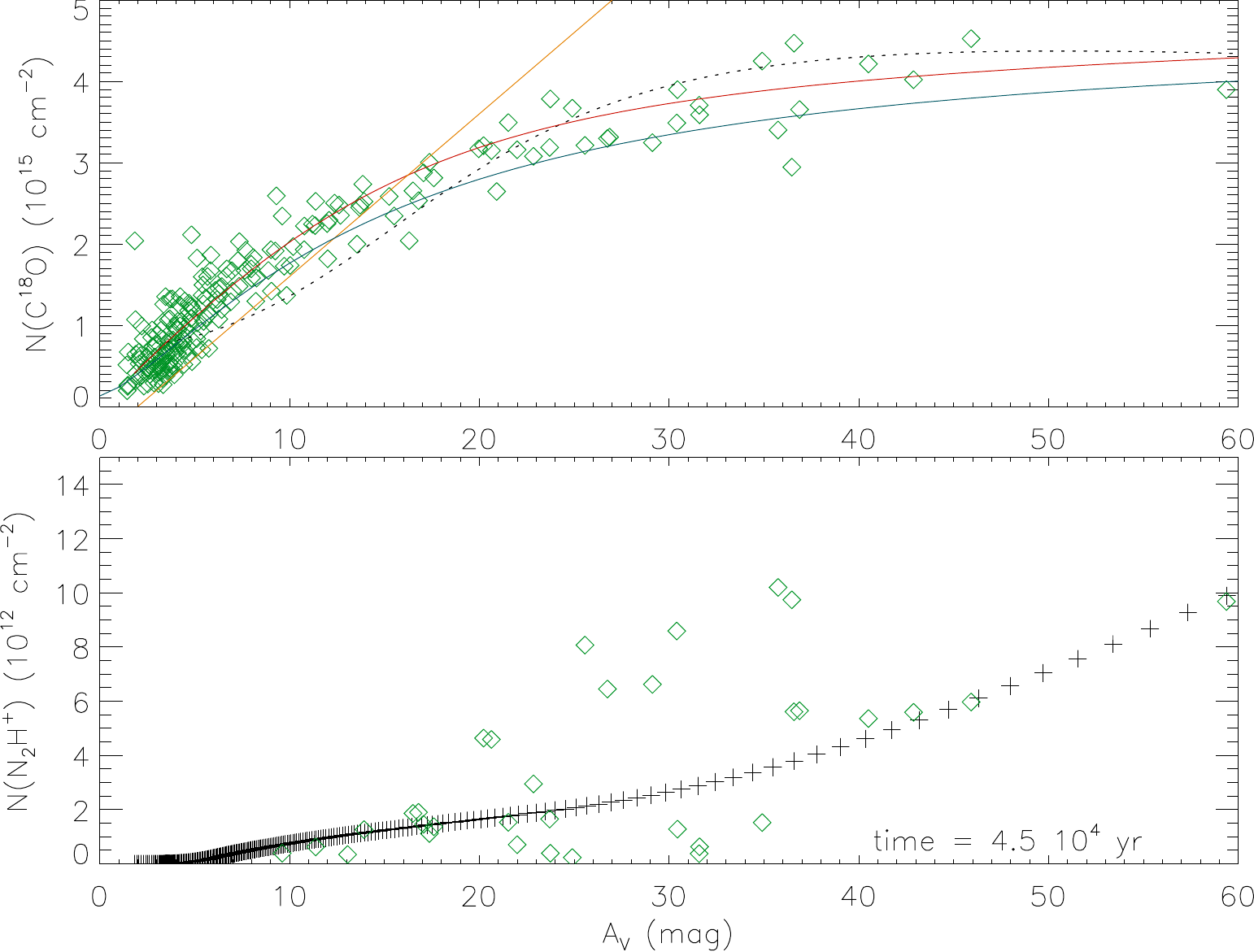}
\caption{Chemical modeling fits to SL42 observations. Green diamonds correspond 
to observed data. The dashed black line (top panel) and black crosses (bottom panel) 
show the best fits of the \protect\citet{2012A&A...543A..38S} model, 
while the red line shows the best fit from the 
\protect\citet{2008ApJ...683..238K} model. The sigmoidal function fitted to the \ceio\ 
data is shown in blue, and the empirical fit to undepleted clouds 
\protect\citep{2006A&A...447..597K} is in orange.}
\label{figure:model}
\end{figure}

The plot of $N(\ceio)$ versus \av\ based on our observations and
extinction estimates data from 2MASS and \textit{Herschel} is shown in Figure~\ref{fig:avco}.
The straight line is the linear relation from \citet{2006A&A...447..597K},
and the clear divergence of the data from this line at high extinctions
($\av \ga 15$) is most likely caused by depletion. No data on ice-phase abundances 
are available that would allow us to directly quantify the uptake of CO into grain
mantles in SL42, but we note that the form of the variation of
$N(\ceio)$ with \av\ in Figure~\ref{fig:avco} is qualitatively very similar to that
found by \citet{2010ApJ...720..259W} for Taurus. This variation can be
described by a sigmoidal function of the form $y = a_{2} + (a_{1}-a_{2})/(1 + (x/x_{0})^{p})$. 
A fit of this form to our data for SL42 is shown in 
Figure~\ref{fig:avco} and compared with the corresponding curve for Taurus from
\citet{2010ApJ...720..259W}. We conclude that the $N(\ceio)$ versus \av\
relation in SL42 follows the same general empirical law as that found for 
Taurus, but with lower levels of depletion.

As previously mentioned, the \textit{Herschel} \av\ data in Figure~\ref{fig:avco} were scaled 
in order to match 2MASS data at low extinctions. The scaling was uniform and made under the 
assumption that the dust opacity ($\kappa_{250\mu\rm{m}}$) remains constant throughout the cloud. 
\citet{2012ApJ...751...28M} found that this assumption is not necessarily true; in the dense molecular 
interstellar medium of the galactic plane, the dust opacity increases with decreasing temperature. 
Because our dust temperature maps are averaged along the line of sight, it is not possible to correct 
for the likely increasing dust opacity at the core of SL42. However, the dust opacity should vary 
at most by a factor of $2-3$ times from edge to core of SL42, and this would 
effectively compress Figure~\ref{fig:avco} along the $x$-axis, only at the highest extinctions. The 
general conclusion of CO depletion at the core of SL42 would be unaffected.

The offsets in peak \ceio\ and \diaz\ concentrations provide supporting evidence for CO depletion in 
the core of SL42. Offsets between \ceio\ and \diaz\ are seen in many other starless cores, e.g., L1498, 
L1495, L1400K, L1517B, L1544 \citep{2002ApJ...569..815T}, and Barnard 68 \citep{2003ApJ...586..286L}. 
The differentiation of \diaz\ and CO originates from the main production pathway of \diaz\ in the 
reaction between ${\rm N_2}$ and ${\rm H_{3}^{+}}$. The ${\rm N_2}$ molecule is formed from 
relatively slow neutral-neutral reactions, and its abundance builds up later than that of CO. On the 
other hand, the ${\rm H_{3}^{+}}$ ion increases strongly only after the depletion of CO in the densest 
parts of molecular clouds. The different behaviors of C-bearing and N-bearing species have been 
discussed, e.g., by \citet{2005A&A...436..933F} and by \citet{2012A&A...541A..32F}, see their Appendix A. 
At very late stages of chemical evolution \diaz\ is also likely to become 
depleted, but previous observations suggest that even then the \diaz$/$CO abundance ratio remains 
high. One possible explanation suggested by \citet{2005A&A...436..933F} is that a low sticking coefficient 
of atomic N could contribute to the availability of ${\rm N_2}$ also in highly depleted regions. 

\subsection{Dynamics}

The Plummer-like density profile with three parameters, $n_0$ (the
central density), $r_0$ (the characteristic radius of the inner region
with a flat density distribution), and $\eta$ (describing the density
power-law far from the center) has been used in several previous
studies for fitting the observed column density profiles of prestellar
cores or starless filaments \citep{2001ApJ...547..317W,2002ApJ...569..815T}. 
Dust continuum maps
have often suggested very steep ($\eta \geq 4$) density gradients near
the edges of prestellar cores \citep{2001ApJ...547..317W}, but in
the case of SL42 the azimuthally averaged column density profile can
be fit with $\eta=2$ up to a radial distance where the core is melted into
the envelope of low-column density gas. \citet{2002ApJ...569..815T} found similar
outer profiles as in SL42 for the prestellar cores L1495 and L1400K.

The average density profile obtained for SL42, $n(r) =n_{0}/(1+(r/r_{0})^{2})$, 
can also be used to approximate a Bonnor-Ebert
sphere (BES), which is a pressure-bound, isothermal, hydrostatic
sphere. As evident from Figure~\ref{fig:avprofile}, the structure 
of SL42 differs, however, clearly from a critical BES which has been frequently 
used to describe globules and dense cores (e.g., \citealt{2001Natur.409..159A}). 
The distribution, with a small flat radius $r_0$ compared to
the outer radius of the core corresponds to a highly super-critical
BES with $\xi_{\rm out} \sim 25$ where the balance between forces is
unstable against any increase of the outside pressure (dashed 
green line in Figure~\ref{fig:avprofile}). The overall agreement is, however, better with the adopted 
Plummer-like function with $n(r) \sim r^{-2}$ at large radii.

Clear deviations from the critical BES structure have been 
previously seen in clouds with regular shapes other than SL42. \citet{2005AJ....130.2166K} 
found in their NIR study of Bok globules that while several starless globules 
can be approximated by nearly critical 
BESs with $\xi_{\rm out}=6.5\pm2$, the column density profiles of star forming 
globules resemble those for super-critical BESs. The largest value of 
$\xi_{\rm out}$ was found for the core L694-2 with in-fall features in 
molecular line spectra. The hydrodynamical models of \citet{2005AJ....130.2166K} 
are consistent with these observations: collapsing cores mimic 
super-critical BESs with $\xi_{\rm out}$ increasing with time.

A density distribution approaching the power-law $n(r) \propto r^{-2}$
at large values of $r$ was found in the early numerical simulations of
the collapse of an isothermal sphere by \citet{1968ApJ...152..515B} and 
\citet{1969MNRAS.145..271L}, and the existence of
a self-similar solution with this form was shown by \citet{1969MNRAS.144..425P} 
by analytic methods. This motivated \citet{2010MNRAS.402.1625K} 
to describe the collapse of a prestellar core
by a sequence of BESs where the central density increases and the flat
radius decreases while the outer radius is constant and the density
profile in the outer parts keeps its scaling as $r^{-2}$ (see their
Figure 1). The density distribution approaches that of a singular
isothermal sphere ($n(r) \propto r^{-2}$ throughout) which is the
initial configuration in the inside-out collapse model of \citet{1977ApJ...214..488S}.

\citet{2010MNRAS.402.1625K} predict the flat radius $r_0$ to be equal to 
the product of the sound speed and the free-fall time in the core center
(see their Equation (2)). Allowing for turbulent velocity dispersion we may write
a slightly modified formula for $r_0$:
\begin{equation}
r_{0}=\sqrt{\frac{3\pi (c_{s}^2+\sigma_{\rm{N.T.}}^2)}{32G\rho_0}},
\end{equation}
where $\rho_0$ is the central density. Substituting the adopted
values of the parameters, $T_{\rm{kin}} = 10$ K,
$\sigma_{\rm{N.T.}} = 260$ m s$^{-1}$, $n_0 = 1.2\times10^6$ cm$^{-3}$, we
get for the expected flat radius $r_0 = 1900$ AU or 15\arcsec, i.e.,
the same value as determined from the observations.

The calculated density profile and the agreement between the flat radius
with the theoretical prediction suggest that SL42 is indeed contracting.
According to the model of \citet{2010MNRAS.402.1625K} the contraction speed
is highest around the flat radius (15\arcsec\ in our case) and clearly
subsonic at this stage of the evolution. The detection of infall
is not possible with the resolution and sensitivity of the present data.

\subsection{Modeling}
\label{sec:discuss_modeling}

Our pseudo-time dependent chemistry model discussed in Sections~\ref{sec:chem_model} 
and \ref{sec:spec_line_model} failed to reproduce the \ceio\ column densities in the outer 
parts of SL42. Judging from the deduced density profile, the cloud is evolving dynamically, 
and probably the main difficulty here is to find realistic initial conditions for the chemistry 
taking the earlier history of the cloud into account. In their genuinely time-dependent model 
for a contracting core, \citet{2010MNRAS.402.1625K} found that CO reaches its equilibrium abundance 
in a short time compared with the dynamical timescale. It seems therefore justified to 
examine if a steady-state model can explain the observed \ceio\ column density profile.

Assuming a steady state, the fractional \ceio\ abundance can
be described by a simple function of the gas density. The accretion
and desorption rates (${\rm cm^{-3}~s^{-1}}$) in an isothermal cloud can be written as 
$A n(\h2) n(\ceio,{\rm gas}) = A n(\h2)^2 X(\ceio,{\rm gas})$ and 
$B n(\ceio,{\rm dust}) = B n(\h2) X(\ceio,{\rm dust})$, respectively. The adsorption coefficient $A$ and
the desorption coefficient $B$ depend on the dust properties,
temperature, and the flux of heating particles (e.g., \citealt{1985A&A...144..147L,
1993MNRAS.263..589H,2002ApJ...565..344C,2002A&A...391..275H,
2008ApJ...683..238K,2010MNRAS.402.1625K}). The
coefficients $A$ and $B$ have the following relationships with the
desorption and depletion timescales, $\tau_{\rm ON}$ and $\tau_{\rm
OFF}$, defined by \citet{2008ApJ...683..238K}: $A n(\h2) = 1/\tau_{\rm
ON}$, $B=1/\tau_{\rm OFF}$. The definitions of $\tau_{\rm ON}$ and
$\tau_{\rm OFF}$ are given in Equations (8) and (11) of \citet{2008ApJ...683..238K}.

In what follows we denote the no-depletion fractional abundance of \ceio\ 
by $X_0$ ($=X(\ceio,{\rm gas})+X(\ceio,{\rm dust})$), and the
gas-phase fractional abundance simply by $X$ ($=X(\ceio,{\rm gas})$).
The steady-state condition, $A n(\h2)^2 X = B n(\h2) (X_0-X)$, implies
the relation (\citealt{2008ApJ...683..238K} Equation (12)):
\begin{equation}
X = \frac{X_0} {1 + \tau_{\rm OFF}/\tau_{\rm ON}}.
\end{equation}
Using the notation of \citet{2002A&A...391..275H}, Equation (8), this equation
becomes $X=X_0/(1 + \frac{A}{B} n({\rm H_2}))$. In the idealized
situation where the gas isothermal, the grain size distribution
remains unchanged, and the same desorption mechanisms are effective in
the whole volume considered, the ratio $A/B$ is constant, and $X$ is a
function of $n({\rm H_2})$ only.

In the steady-state model, several (rather uncertain) parameters affecting the
gas-phase \ceio\ abundance degenerate into
one quantity, the ratio $A/B$ with the dimension ${\rm cm^{-3}}$. This
ratio can be therefore identified with the inverse of the
characteristic density, say $n_{\rm D}$, at which depletion owing to
freezing out becomes substantial. With this definition the equation
for $X$ can be rewritten as
\begin{equation}
X = \frac{X_0} {1 + n({\rm H_2})/n_{\rm D}},
\end{equation}
where $n_{\rm D}$ (using the notation of \citealt{2008ApJ...683..238K}) is
\begin{equation}
n_{\rm D} = \frac{3\times10^{-7}\zeta_{\rm{H_{2}}}/10^{-17}\rm{exp}(-E_{\rm CO}/70)}{S_0 R_{\rm dg}\sigma V_T},
\end{equation}
where $\zeta_{H_2}$ is the cosmic ray ionization rate of \h2,
$E_{\rm CO}$ is the binding energy of CO onto ice, $S_0$ is the
sticking coefficient of CO to the dust in a collision, the product
$R_{\rm dg} \sigma$ is the grain surface area per \h2\ molecule,
and $V_T$ is the thermal speed of \ceio\ molecules.

Assuming an isothermal cloud at 10 K, and using the dust parameters
and the cosmic ray ionization rate listed in \citet{2008ApJ...683..238K},
$n_{\rm D}$ obtains the value $\sim 1.2\times10^{4}$ cm$^{-3}$. Here the
assumed total grain surface area is $\sim 1.4\times10^{-21}$ cm$^{2}$ (per
\h2) and $\zeta_{\rm H_2} = 3\times10^{-17}$ s$^{-1}$. The density
$n_{\rm D}$ can be made higher by, e.g., increasing the minimum grain
size (which decreases the total grain surface area and reduces
accretion) or by increasing the cosmic ray ionization rate (which
enhances desorption).

We have plotted in Figure \ref{figure:model} the \ceio\ column density as a 
function of \av\ as predicted by the steady-state model (red solid line), together with
the $N(\ceio)$ versus \av\ from observations (crosses) and the
profile predicted by our time-dependent chemistry model (black dashed
line) which were discussed in Section~\ref{sec:chem_model}. The \av\ axis of the model data
(lines) is calculated from the spherically symmetric cloud model
described in Section~\ref{section:MassDynamics}. For the steady-state depletion model a rough
agreement with the observed column density profile was attained by
setting $X_0 (\ceio)=2.4\times10^{-7}$, $n_{\rm D}=3.8\times10^{4}$~cm$^3$. 
The change in $n_{\rm D}$ with respect to the value quoted above
corresponds to a decrease in the total grain surface area by a factor
of three, or to an increase in the cosmic ray ionization rate (or the
overall efficiency of desorption mechanisms) by the same amount. A
decrease in the effective total grain surface area can be justified by
assuming that CO is efficiently desorbed from the smallest grains by
other processes than cosmic ray heating (\citealt{2002A&A...391..275H},
Section~\ref{sec:summary}). We note that \citet{2010MNRAS.402.1625K} needed to increase the
desorption rate by an order of magnitude from that implied by the
direct cosmic ray heating (denominator of the formula for $n_{\rm D}$)
to obtain agreement with the observed CO spectra toward L1544. 

The prediction of the steady-state model agrees better 
with the observed CO column density profile than the time-dependent model. 
We note that while the time-dependent model was optimized to 
match with the observed \ceio/\diaz\ column density toward the core 
center, we have no estimate for the \diaz\ column density profile based 
on the steady-state model, hence we cannot in this case attempt to 
simultaneously match \ceio\ and \diaz\ profiles to observations. Thus, we 
cannot draw conclusions as to which model is better in the sense of 
reproducing all of the observed data.

The agreement between \ceio\ observations and the 
steady-state model is not perfect as the model column density does not 
seem to rise steeply enough. Furthermore, the observations and the model 
predictions at low extinctions appear to have a horizontal offset of about 
2 magnitudes from the empirical linear relationship of \citet{2006A&A...447..597K}. 
This offset can probably be accounted for by the 
uncertainty related to the zero levels of the pipeline reduced 
\textit{Herschel} maps. The calibration can be improved in the future when the 
\textit{Planck} data of the region become publicly available.

The prediction of the steady-state model gives equally good agreement
with the observed \ceio\ column density profile as the sigmoidal
function. This suggests the characteristic shape of the $N(\ceio)$ versus $A_{\rm V}$ 
correlation seen in SL42 and previously in Taurus reflects the density
structure in the core envelopes. In the steady-state model, local
variation in the CO depletion depends mainly on the density.

Both the density structure and the CO depletion law found in
SL42 agree with the model of \citet{2010MNRAS.402.1625K} for a thermally
super-critical prestellar core. The outer density profile follows the
power law $n(r) \propto r^{-2}$, the flat radius is consistent with
the estimated physical parameters in the core center, and the depletion
time scale is short compared with the contraction time scale so that
the CO abundance distribution does not deviate much from that for a
static cloud with the same physical characteristics at any moment.


\section{Summary}
\label{sec:summary}

We have studied the structure and chemistry of the relatively isolated
prestellar dense cloud Sandqvist \& Lindroos 42 in
Corona Australis using the dust emission maps observed with the
\textit{Herschel} satellite and \ceio\ and \diaz\ line maps observed with the SEST
radio telescope. The \h2\ column density map as calculated from \textit{Herschel} data 
shows a sharp peak suggesting a high central density. The azimuthally
averaged column density distribution can be fitted with the
characteristic prestellar density profile which scales as $r^{-2}$ in
the outer part.

$\rm{C^{18}O}$~and \diaz\ peak at different locations, about 50\arcsec\ 
(6500 AU) apart. The \diaz\ distribution is compact, and the maximum coincides with
the $N(\h2)$ peak. The separation between the \diaz\ and \ceio\ can be
probably understood in terms of CO depletion. According to chemistry
models, the \diaz\ abundance builds up slowly and is benefited by the
disappearance of CO in the densest parts of molecular clouds.

The relation between $N(\ceio)$ and \av\ (or $N(\h2)$) is clearly not linear.
The curved distribution of points on a $N(\ceio)$ versus \av\ plot is similar
to that found by \citet{2010ApJ...720..259W} for Taurus and can be fitted with a
sigmoidal function. The curvature is most likely caused by accretion of
CO onto grain surfaces.

The shape of the $N(\ceio)$ versus \av\ relation can be explained fairly well
with a steady-state depletion model, where the fractional \ceio\ abundance
is a simple function of density: $X(\ceio)\sim X_0(\ceio)/(1+n/n_{\rm D})$,
where $X_0(\ceio)$ is the no-depletion fractional abundance
and $n_{\rm D}$ is a characteristic density depending on the
desorption and adsorption rates \citep{2008ApJ...683..238K,2002A&A...391..275H}.

The density structure and the CO depletion law observed in SL42 agree
with the models of \citet{2008ApJ...683..238K,2010MNRAS.402.1625K} for the evolution of
a thermally super-critical prestellar core, where the subsonic contraction
is described by a series of nearly hydrostatic cores with decreasing
flat radii and $r^{-2}$ density profiles in the outer parts. Our molecular 
line data are not appropriate for detecting the possible radial inflows.


\section*{Acknowledgements}

\emph{Financial support for this research was provided by the NASA Exobiology and 
Evolutionary Biology program (grant NNX07AK38G), the NASA Astrobiology Institute 
(grant NNA09DA80A), the NASA New York Space Grant Consortium, the Academy of 
Finland (grants 132291, 250741, 73727, and 74854), and the Magnus Ehrnrooth 
Foundation of the Finnish Society of Sciences and Letters. This research has made 
use of data from the Two Micron All Sky Survey, which is a joint project of the 
University of Massachusetts and the Infrared Processing and Analysis Center, funded 
by NASA and the National Science Foundation.}

\bibliography{SL42refs}

\clearpage


\clearpage


\clearpage



\clearpage

\clearpage



\begin{table}
\caption{Line Characteristics and the \ceio\ and \diaz\ Column Densities in Selected Positions}
\begin{tabular}{rcccccr}\hline
\multicolumn{6}{c}{\ceio\ $J=2-1$} \\
\hline
($\Delta\alpha,\Delta\delta$)&$T_{A,\rm{peak}}^*$&$v_{\rm{LSR}}$&$\Delta v$&$\tau$&$T_{ex}$& $N(\ceio)$\\
\multicolumn{1}{c}{(\arcsec)} & (K) & (\kms) & (\kms)& & (K) &($10^{15}$~cm$^{-2}$)\\
\hline
   (0,0)   & 3.52 (0.07) & 5.53 (0.01) & 0.72 (0.01) & $\ldots$ & 6.65 (0.28) & 4.53 (0.09) \\
  ($-$25,34) & 3.66 (0.07) & 5.45 (0.01) & 0.70 (0.01) & $\ldots$ & 6.77 (0.39) & 4.47 (0.11) \\  
 (34,25)  & 3.17 (0.06) & 5.48 (0.01) & 0.78 (0.01) & $\ldots$ & 6.87 (0.38) & 4.18 (0.09) \\
   (59,$-$9) & 3.01 (0.06) & 5.52 (0.01) & 0.77 (0.01) & $\ldots$ & 7.40 (0.41) & 3.51 (0.06) \\ 
\hline
\multicolumn{6}{c}{\diaz\ $J=1-0$} \\ 
\hline
($\Delta\alpha,\Delta\delta$)&$T_{A,\rm{peak}}^*$&$v_{\rm{LSR}}$&$\Delta v$&$\tau$&$T_{ex}$& $N(\diaz)$\\
\multicolumn{1}{c}{(\arcsec)} & (K) & (\kms) & (\kms) & & (K) & ($10^{12}$~cm$^{-2}$)\\
\hline
  (0,0)    &  0.63 (0.07) & 5.67 (0.02) & 0.52 (0.04) & 4.75 (1.98) & 4.14 (0.32) & 5.97 (2.62) \\
  ($-$25,34) &  0.51 (0.06) & 5.50 (0.02) & 0.73 (0.05) & $\ldots$ & 5$^a$     & 2.35 (0.45) \\
  (34,25)  &  1.28 (0.05) & 5.66 (0.01) & 0.52 (0.01) & 8.00 (0.89) & 4.98 (0.11) &13.40 (1.60) \\
(59,$-$9) &  0.99 (0.06) & 5.70 (0.01) & 0.50 (0.02) & 9.03 (1.36) & 4.29 (0.11) &11.60 (1.87)  \\
 \hline
\end{tabular}
\label{table:lineparameters}
\vspace{3mm}
\textbf{Note.} $^a$ $\tau$ and $T_{ex}$ could not determined. The \diaz\ column density was calculated by assuming optically thin emission with $T_{ex}=5$.
\end{table}

\clearpage

	\clearpage

	\clearpage


	\clearpage



\clearpage


\clearpage

\end{document}